\documentclass[preprint, 12pt, authoryear]{elsarticle}

\usepackage[margin=3cm]{geometry}
\usepackage{graphicx}
\usepackage{amssymb}
\usepackage{amsmath}
\usepackage{url}
\usepackage{pdflscape}

\newcommand{\lppls}{\mbox{LPPLS}}
\newcommand{\sse}{\mbox{SSE}}

\makeatletter
\def\ps@pprintTitle{%
  \let\@oddhead\@empty
  \let\@evenhead\@empty
  \let\@oddfoot\@empty
  \let\@evenfoot\@oddfoot
}
\makeatother

\usepackage{etoolbox}
\patchcmd{\MaketitleBox}{\footnotesize\itshape\elsaddress\par\vskip36pt}{\footnotesize\itshape\elsaddress\par\parbox[b][36pt]{\linewidth}{\vfill\hfill\textnormal{\today}\hfill\null\vfill}}{}{}%
\patchcmd{\pprintMaketitle}{\footnotesize\itshape\elsaddress\par\vskip36pt}{\footnotesize\itshape\elsaddress\par\parbox[b][36pt]{\linewidth}{\vfill\hfill\textnormal{\today}\hfill\null\vfill}}{}{}%

\begin{document}

\begin{frontmatter}

\title{Modified Profile Likelihood Inference and Interval Forecast of the Burst of Financial Bubbles}

\author[1,2]{Vladimir Filimonov}
\ead{vfilimonov@ethz.ch}

\author[1]{Guilherme Demos}
\ead{gdemos@student.ethz.ch}

\author[1,3]{Didier Sornette}
\ead{dsornette@ethz.ch}

\address[1]{Dept. of Management, Technology and Economics, ETH Z\"{u}rich, Z\"{u}rich, Switzerland}
\address[2]{Department of Economics, Perm State University, Perm, Russia}
\address[3]{Swiss Finance Institute, c/o University of Geneva}

\begin{abstract}
{   We present a detailed methodological study of the application of the modified profile likelihood method 
for the calibration of nonlinear financial models characterised by a large number of parameters.
We apply the general approach to the Log-Periodic Power Law Singularity 
(LPPLS) model of financial bubbles. This model is particularly relevant because one of its parameters, the critical time $t_c$
signalling the burst of the bubble, is arguably the target of choice for dynamical risk management.
However, previous calibrations of the LPPLS model have shown that the estimation of $t_c$ is in general 
quite unstable. Here, we provide a rigorous likelihood inference approach to determine $t_c$, which takes
into account the impact of the other nonlinear (so-called ``nuisance'') parameters for the correct
adjustment of the uncertainty on $t_c$. This provides a rigorous interval estimation for 
the critical time, rather than a point estimation in previous approaches. As a bonus, the 
interval estimations can also be obtained for the nuisance parameters ($m,\omega$, damping),
which can be used to improve filtering of the calibration results. We show that the use
of the modified profile likelihood method dramatically reduces the number of local extrema by 
constructing much simpler smoother log-likelihood landscapes. The remaining distinct solutions
can be interpreted as genuine scenarios that unfold as the time of the analysis flows, which 
can be compared directly via their likelihood ratio. Finally, we develop a multi-scale profile likelihood
analysis to visualize the structure of the financial data at different scales (typically from 100 to 750 days).
We test the methodology successfully
on synthetic price time series and on three well-known historical financial bubbles. 
}
\end{abstract}



\end{frontmatter}


\section{Introduction}
\label{sec:intro}

Financial bubbles and their subsequent crashes provide arguably the most visible departures from well-functional efficient markets. There is an extensive literature (see e.g. the reviews of \citet{KaizoSor10}\footnote{Long version at http://arXiv.org/abs/0812.2449}, \citet{SornetteZhou2010,BrunnermeierOehmke2013,Xiong2013}) on the causes of bubbles as well as the reasons for bubbles to be sustained over surprising long period of times. One of these views 
emphasises the role of herding behaviour on bubble inflation \citep{SornetteJohansen1999Risk}. When imitation 
is sufficiently strong, a high demand for the asset pushes the price upwards, which itself, and somewhat paradoxically,
increases the demand, propelling further the price upward, and so on, in self-fulfilling positive feedback loops.
In such regimes, the market is mainly driven by sentiment and becomes detached from any underlying economic value. This process is intrinsically unsustainable and the mispricing ends at a critical time, either smoothly (with a correction phase) or abruptly (via a crash). The formulation of this hypothesis of collective herding behavior within rational expectations theory resulted in the so-called Log-Periodic Power-Law Singularity (LPPLS) model, which has been used for many successful ex-post and ex-ante predictions of bubble bursts 
(see e.g. a partial list in~\citep{Sornette2013_JLS}
and a recent implementation for the Chinese bubble and its burst in 2015~\citep{Sornette_etal2015_Shanghai}).

Notwithstanding a number of improvements concerning the calibration of the LPPLS model, including meta-search heuristics
\citep{SornetteZhoufor2006} and reformulation of the equations to reduce the number of nonlinear parameters 
\citep{FilimonovSornette2011_LPPL_calibration}, the calibration of the LPPLS model remains a bottleneck 
towards achieving robust forecasts and a matter of contention \citep{BreeChallet2010,Sornette2013_JLS}. In this context,
the aim of the present paper is to present a fundamental revision of the calibration procedure of the LPPLS model. 
Specifically, we deviate from the traditional ordinary least squares (OLS) calibration that provides \emph{point estimates} of parameters, 
which has been used since the introduction of the model in 1999 \citep{SornetteJohansen1999,SornetteJohansen2000}. 
Instead, we employ a rigorous likelihood approach and, for the first time to the best of our knowledge, we provide \emph{interval estimates} of the parameters, including the most important critical times of market regime changes.

We deliberately avoid dwelling on the derivation of the model and its foundations, and take it as given. We do not discuss 
supporting evidence and critiques of the model, nor address how to apply the LPPLS model
to construct robust signals for extensive backtests or real-time ex-ante predictions. These questions require extensive
analyses and are beyond the scope of the present manuscript. See \citep{SornetteJohansen2010,SornetteZhou2010,Sornette2013_JLS,Sornette_etal2015_Shanghai,QunQunzhididier15}
for investigations in these directions.
 
 The purpose of the present paper is methodological, and the main focus is on the statistical aspects of the theory and 
the corresponding mathematical derivations. One of the major advances of this paper is to formulate the calibration procedure 
so that the critical time $t_c$ is the major parameter of interest in the likelihood inference, while other model parameters 
are treated as so-called \emph{nuisance parameters}. Of course, these other parameters are also 
intrinsic to the model but their existence contributes to the variance of the parameter of key interest. 
Such reformulation of the calibration procedure has its roots in an original idea proposed by~\citet{FilimonovSornette2011_LPPL_calibration}, which was however developed in a crude way and without the proper statistical methodology.

The problem of dealing with nuisance parameters and of quantifying their impact on the uncertainty of the parameter of interest is 
not new in Statistics. However, to our knowledge, it has not been elaborated before in quantitative finance. Frequentist and Bayesian statistical schools have different views on this problem. The main debate between the supporters of likelihood-based versus Bayesian approaches is whether one should maximize over nuisance parameters (such as in simple profile likelihood) or integrate them out. Both approaches have their pros and cons. In general, the method of profile likelihood is known to often provide biased estimations. However, use of the Bayesian (or integration) approach requires specification of the prior distribution of the parameters, which leads to an extra uncertainty in inference.
Under certain conditions, when the full likelihood function has a complex structure, the two methods can lead to dramatically different estimations~\citep{SmithNaylor1987,Berger1999}. 

We will base our approach on the so-called modified profile likelihood proposed by~\citet{BarndorffNielsen1983} as a higher-order approximation to either a marginal or conditional likelihood function. Being unable to calculate 
the modified profile likelihood exactly due to strong model nonlinearity, we will employ the  approximation suggested by~\citet{Severini1998}, which is equivalent to the exact form up to errors of order $\mathcal{O}(n^{-1})$ for moderate deviations and of order $\mathcal{O}(n^{-1/2})$ in the large deviation sense, where $n$ is the number of data points.
The advantage of this method is that it takes the middle ground in the maximization-vs-integration debates: as shown by \cite{Severini2007}, the modified profile likelihood arises naturally from a non-Bayesian inference with an integrated likelihood and could even be considered as an approximation to a certain class of integrated likelihood functions. At the same time, it does not require specifying a prior density of the nuisance parameters, which makes it perfectly suitable in our case.

In the following, we will guide the reader from the well known OLS calibration procedure and its formulation as a likelihood problem, to the lesser known ``profile likelihood'' and then to the ``modified profile likelihood'', which has been essentially ignored in the applied literature.
The modified profile likelihood allows one to improve the likelihood inference by accounting for the uncertainty of the nuisance parameters.
Having a strong methodological emphasis, we will discuss all concepts and, more important, their assumptions and limitations in all necessary details. While this paper focuses on the LPPLS model, our general presentation and its specific implementation
on the LPPLS model makes it useful as a general guide for
likelihood inference in many other models of quantitative finance.

The paper is organized as follows. Section~\ref{sec:lppls} presents the Log-Periodic Power Law Singularity model and discusses its structure and constraints. Section \ref{sec:ols} presents the Ordinary Least Squares (OLS) method that has been used until now as the standard calibration tool of the LPPLS model, in particular for the estimation of the critical time $t_c$ of the end of the bubble.  
Section \ref{sec:likelihood} introduces the Likelihood and Profile Likelihood approaches.  Section \ref{sec:modified}
presents the general concept of the modified profile likelihood and provides 
a very useful approximated expression for it. 
Parameter estimation uncertainties and the corresponding likelihood intervals are then derived. 
Section \ref{sec:m_w} applies the modified likelihood profile to estimate confidence intervals of the nuisance 
parameters $m$ and $\omega$ as well as the damping variable.
Section~\ref{sec:examples} presents the method of aggregation of the calibrations from different scales and 
illustrates the whole methodology on synthetic price time series. This section ends with the application 
of the method on three well-known historical financial bubbles. Section~\ref{sec:conclusion} concludes.

\section{Log-Periodic Power Law Singularity model}\label{sec:lppls}

The LPPLS model is based on the standard jump-diffusion model, where the logarithm of the asset price $p(t)$ follows a random walk with a varying drift $ \mu(t)$ in the presence of discrete discontinuous jumps:
\begin{equation}\label{eq:JLS}
	\frac {dp}{p} = \mu(t)dt+\sigma(t)dW-\kappa dj.
\end{equation}
Here, $\sigma(t)$ denotes the volatility, $dW$ is the infinitesimal 
increment of a standard Wiener process and $dj$ represents a discontinuous jump such as $j=\chi(t-t_c)$, where $\chi(\cdot)$ is a Heaviside function and $t_c$ denotes the time of the jump. Within the ``bubble-crash'' framework, $t_c$ defines the ``critical time'', which is defined within the rational expectations framework as the most probable time for the crash or change of regime to occur. The parameter $\kappa$ then quantifies the amplitude of the crash when it occurs. The expected value of $dj$ defines the crash hazard rate $h(t)$:  $\text{E}[dj]=h(t)dt$. 

According to the Johansen-Ledoit-Sornette (JLS) model~\citep{SornetteJohansen1999Risk,SornetteJohansen1999,SornetteJohansen2000}, the complex actions of noise traders can be aggregated into the following dynamics of the hazard rate:
\begin{equation}\label{eq:hazard}
	h(t)=\alpha(t_c-t)^{m-1}\big(1+\beta\cos(\omega\ln(t_c-t)-\phi')\big),
\end{equation}
where $\alpha$, $\beta$, $\omega$ and $\phi'$ are some parameters. The core of the model is the singular power law behavior $(t_c-t)^{m-1}$ that embodies the mechanism of the positive feedback at the origin of the formation of bubble leading to a super-exponential price growth. The oscillatory dressing $1+\beta\cos(\omega\ln(t_c-t)-\phi')$ takes into account the existence of a possible hierarchical cascade of panic acceleration punctuating the course of the bubble. The particular form of the log-periodic function $\cos(\omega\ln(t_c-t)-\phi')$ in~\eqref{eq:hazard} is a first-order expansion of the general class of Weierstrass-type functions~\citep{Sornette2002fractal_functions,ZhouSorRen03} that describes the discrete-scale invariance around tipping points in complex natural and socio-economic systems~\citep{Sornette1998_discrete_scales,Sorpredic02}.

Under the no-arbitrage condition ($\text{E}[dp]=0$), the excess return $\mu(t)$ is proportional to the crash hazard rate $h(t)$: $\mu(t) = \kappa h(t)$. Then direct solution of the equation~\eqref{eq:JLS} with the given dynamics of the hazard rate~\eqref{eq:hazard} under the condition that no crash has yet occurred ($dj=0$) leads to the following Log-Periodic Power Law Singularity (LPPLS) equation for the expected value of a log-price:
\begin{equation}\label{eq:lppl}
	 \lppls(t)\equiv\text{E}[ \ln p(t)]=
	 A+B(t_c-t)^m+C(t_c-t)^m\cos(\omega\ln(t_c-t)-\phi),
\end{equation}
where $B=-\kappa\alpha/m$ and $C=-\kappa\alpha\beta/\sqrt{m^2+\omega^2}$. It is important to stress that the exact solution~\eqref{eq:lppl} describes the dynamics of the average log-price only up to critical time $t_c$ and cannot be used beyond it. This critical time $t_c$ corresponds to the termination of the bubble and indicates the change to another regime, which could be a large crash or a change of the average growth rate. Nevertheless, in practical applications, one often heuristically extends~\eqref{eq:lppl} for $t>t_c$, assuming the validity
of a time-inversion symmetry of the price trajectory around $t_c$ \citep{JohansenSornette1999,ZhouSorRen03}:
\begin{equation}\label{eq:lppl_symmetry}
	\lppls(t)=A+B|t_c-t|^m+C|t_c-t|^m\cos(\omega\ln|t_c-t|-\phi).
\end{equation}

The LPPLS model in its original form~\eqref{eq:lppl} or~\eqref{eq:lppl_symmetry} is described by three linear parameters ($A,B,C$) and four nonlinear parameters ($m,\omega,t_c,\phi$). As discussed further in Section~\ref{sec:ols}, the calibration of the model can be performed using a two-stage procedure. First, the linear parameters $A,B,C$ for fixed values of $m,\omega,t_c,\phi$ can be obtained directly via the solution of a matrix equation. Second, the non-linear parameters $m,\omega,t_c,\phi$ can be found using a nonlinear optimization method. 
Notwithstanding the reduction from 7 to 4 of the number of parameters to determine, the numerical optimization is not straightforward, as the cost function possesses a quasi-periodic structure with many local minima. Any local optimization algorithm fails here and an extra layer involving so-called metaheuristic algorithms~\citep{Talbi2009Metaheuristics} is needed in order to find the global optimum. 
Thus, the \emph{taboo search}~\citep{Cvijovicacute1995} has often been used to perform this metaheuristic determination
of the 4 nonlinear parameters of the LPPLS function~\eqref{eq:lppl}.

A better approach has been suggested by
\citet{FilimonovSornette2011_LPPL_calibration}, which consists in reformulating the model~\eqref{eq:lppl} 
in a way that significantly simplifies the calibration procedure. The reformulation is based on the variable change 
\begin{equation}\label{eq:C12}
	C_1=C\cos\phi,	\quad 	
	C_2=C\sin\phi,
\end{equation}
so that equation~\eqref{eq:lppl} becomes
\begin{equation}\label{eq:lppl2}
	\lppls(t)=A+B|t_c-t|^m+C_1|t_c-t|^m\cos(\omega\ln|t_c-t|)+C_2|t_c-t|^m\sin(\omega\ln|t_c-t|).
\end{equation}
In this form, the LPPLS function has only 3 nonlinear ($t_c, \omega, m$) and 4 linear $A,B,C_1,C_2$ parameters. As shown in~\citep{FilimonovSornette2011_LPPL_calibration}, this transformation significantly decreases the complexity of the fitting procedure and improves its stability tremendously. This is because the modified cost function for~\eqref{eq:lppl2} is now free from quasi-periodicity and enjoys good smooth properties with one or a few local minima in the case where the model is appropriate to the empirical data. 


Let us complement this exposition by briefly discussing the constraints in parameter space. Since the integral 
$\int_{t_0}^{t_c}h(t)dt$ of the hazard rate~\eqref{eq:hazard} gives the probability of the occurrence of a crash, 
it should be bounded by $1$, which yields the condition $m<1$. 
At the same time, the log-price~\eqref{eq:lppl} should also remain finite for any $t\leq t_c$, which implies the condition $m>0$. In addition, 
in order for the LPPLS formula to capture the super-exponential acceleration associated with a bubble, we need  $B<0$. 
Finally, the hazard rate $h(t)$ is non-negative by definition \citep{Bothmer2003}, which translates into 
the constraint $D={m|B|}/{\omega|C|}>1$, where $D$ is the so-called \emph{damping} parameter.

Additional constraints have been proposed, based on compilations of extensive analyses of historical bubbles~\citep{SornetteJohansen2001,SornetteJohansen2010,SornetteLin2009}. 
\citet{SornetteJohansen2010} document approximate
Gaussian distributions of $\omega$ and $m$ with the corresponding mean and standard 
deviations: $\omega\approx6.35\pm1.55$ and $m\approx0.33\pm0.18$. In practical implementations, these constraints are slightly modified in order to minimize errors of type I (incorrect rejection of the LPPLS hypothesis). In particular, the constraints for $\omega$ are often pushed upward to avoid small angular log-frequencies that can spuriously appear as a result of improper fitting of trends. Finally, the strict theoretical constraint $D>1$ on the damping parameter is derived under the assumption that the crash occurs in one immediate negative jump.
As this is in general counterfactual (a crash has usually a duration of weeks to months, and is characterized by a large drawdown \citep{SornetteJohansenoutlier01,SornetteJohansen2010}), 
the constraint $D>1$ can be relaxed~\citep{Sornette_etal2015_Shanghai}. To sum up, the following set of theoretical and empirical constraints on the parameters can be regarded as the stylized features of LPPLS:
\begin{equation} \label{eq:constrains}
    0.1\lessapprox m\lessapprox0.9,\quad 6\lessapprox\omega\lessapprox13,\quad B<0,\quad D=\frac{m|B|}{\omega\sqrt{C_2^2+C_2^2}}\gtrapprox0.8.
\end{equation}

The last important question we need to address is the ``nature'' of the proper time $t$ to consider. The standard jump-diffusion equations~\eqref{eq:JLS} can be applied either to \emph{calendar time}, where $t$ continuously increases, or to \emph{business time}, where weekends and non-trading days are omitted within a discrete version where $t$ increases by 1 day from Friday to the next Monday. The standard discrete normal diffusion process (i.e. without the jump term) is invariant to this transformation of the calendar, assuming that the drift $\mu$ and volatility $\sigma$ are rescaled properly. While the LPPLS model~\eqref{eq:lppl} is not invariant to such time change, both approaches are possible. In previous works, the calibration of the model has been performed using calendar time, assuming that the price variations over non-trading dates are non-observable but nevertheless embody an information flow that impacts the overall price dynamics. Here, we also use calendar time.

\section{Nonlinear regression and Ordinary Least Squares fitting \label{sec:ols}}

\subsection{Solution of the nonlinear regression problem}

In the LPPLS framework, forecasting the termination of a bubble amounts to finding the best estimation of the critical time $\hat t_c$. This requires calibrating formula~\eqref{eq:lppl2} on the observed price trajectory in order to determine $t_c$ together with all the other parameters of the model, ${\psi=\{ m, \omega, A, B, C_1, C_2\}}$. In previous works, this was done via a nonlinear regression of the vector of log-prices $Y=\{\ln p(\tau_i)\}$ on the vector of observation dates $X=\{\tau_i: \tau_i\in[t_1,t_2]\}$, where $[t_1,t_2]$ denotes the window of analysis. The \emph{Ordinary Least Squares (OLS)} method amounts to minimizing the sum of squared residuals ($\varepsilon(\tau_i;t_c,\psi)=\ln p(\tau_i)-\lppls(\tau_i;t_c,\psi)$) between $Y$ and the $\lppls$ formula,
\begin{equation}\label{eq:OLS}
        \{\hat t_c, \hat\psi\} = \arg\min_{t_c,\psi}\sse(t_c,\psi),
\end{equation}
where the sum of squared errors (SSE) is given by
\begin{equation}\label{eq:SSE}
	\sse(t_c,\psi)=\sum_{i=1}^n\big(\varepsilon(\tau_i;t_c,\psi)\big)^2
	\equiv\sum_{i=1}^n\big(\ln p(\tau_i)-\lppls(\tau_i;t_c,\psi)\big)^2~.
\end{equation}
Minimization of such nonlinear multivariate cost function is a highly non-trivial task due to presence of multiple local minima, where the local optimization algorithm can get trapped. 

In the original formulation of the model~\eqref{eq:lppl}, three linear parameters $A, B, C$ can be slaved to the nonlinear 
parameters $t_c,m,\omega,\phi$~\citep{SornetteJohansen1999Risk}. This decreases significantly the complexity
of the calibration problem, but does not remove the quasi-periodic structure of the cost-function with many minima.
As mentioned above, this requires metaheuristic methods for the optimization. In the reformulated model~\eqref{eq:lppl2},
the complexity of the optimization problem is further decreased by transforming the non-linear phase 
into a linear parameter. Then, one can first slave the four linear parameters $A,B,C_1,C_2$ to the three 
remaining nonlinear parameters $t_c,m,\omega$ \citep{FilimonovSornette2011_LPPL_calibration}. 
The minimization problem~\eqref{eq:OLS} is thus transformed into:
\begin{equation}\label{eq:S_arg_new}
	\{\hat t_c,\hat m,\hat \omega\}=\arg\min_{t_c,m,\omega}F_1(t_c,m,\omega),
\end{equation}
where the cost function $F_1(t_c,m,\omega)$ is given by
\begin{equation}\label{eq:S1_new}
	F_1(t_c,m,\omega)= \min_{A,B,C_1, C_2}
	\sse(t_c,m,\omega,A,B,C_1,C_2)~.
\end{equation}
The optimization problem in~\eqref{eq:S1_new} has a unique solution obtained directly from the first-order condition:
\begin{equation}\label{eq:ABC12}
  \left(\begin{array}{cccc}
    n & \sum f_i & \sum g_i & \sum h_i \\
    \sum f_i & \sum f_i^2 & \sum f_ig_i & \sum f_ih_i\\
    \sum g_i & \sum f_ig_i & \sum g_i^2  & \sum g_ih_i\\
    \sum h_i & \sum f_ih_i & \sum g_ih_i  & \sum h_i^2\\
  \end{array}\right)\left(\begin{array}{c}
    \hat A\\ \hat B\\ \hat C_1 \\ \hat C_2  
  \end{array}\right)=
  \left(\begin{array}{c} 
    \sum y_i \\ \sum y_if_i \\ \sum y_ig_i \\ \sum y_ih_i
  \end{array}\right)
\end{equation}
where 
\begin{align}\label{eq:yfgh}
	y_i&=\ln p(\tau_i),\nonumber\\
	f_i&=|t_c-\tau_i|^m,\nonumber\\
	g_i&=|t_c-\tau_i|^m\cos(\omega\ln|t_c-\tau_i|),\\
	h_i&=|t_c-\tau_i|^m\sin(\omega\ln|t_c-\tau_i|).\nonumber
\end{align}

As discussed in Section~\ref{sec:lppls}, the reduction from 4 to 3 nonlinear parameters decreases dramatically the number of local extrema to only a few, so there is not much need for metaheuristic methods such as the Taboo search~\citep{Cvijovicacute1995}, which
was previously the main tool of the LPPLS calibration \citep{SornetteJohansen2000}. In most cases, a single ``quasi-local'' optimization algorithm such as the Nelder-Mead simplex method~\citep{NelderMead1965} can reliably find the absolute minimum of 
$F_1(t_c,m,\omega)$ given by expression (\ref{eq:S1_new}).
In complicated cases the Nelder-Mead simplex method
can be complemented by employing \emph{repeated local searches}.
This amounts to start the local search routine from multiple different initial points and then select the best solution.

\subsection{Estimation of the critical time}

In practical applications, the calibration of the LPPLS model often aims at forecasting the critical time $t_c$, 
because it is, by construction of the LPPLS model, the end of the bubble regime. This suggests
to develop a special treatment for $t_c$. In this spirit, \citet{FilimonovSornette2011_LPPL_calibration} 
suggested to reformulate the optimization problem~\eqref{eq:S_arg_new} by subordinating the logperiodic
angular frequency $\omega$ and power law exponent $m$ to $t_c$:
\begin{equation}\label{eq:S_arg2_new}
	\hat t_c=\displaystyle\arg\min_{t_c} F_2(t_c),
\end{equation}
where 
\begin{equation}\label{eq:S2_new}
	F_2(t_c)=\displaystyle\min_{\omega,m} F_1(t_c,m,\omega),
	\quad
	\{\hat m(t_c),\hat \omega(t_c)\}=\arg\min_{m,\omega}F_1(t_c,m,\omega)
\end{equation}
and $F_1(t_c, m, \omega)$ is given by~\eqref{eq:S1_new}. 

In general, such extra subordination dramatically reduces the number of local extrema of the cost-function. As we will see later from Figure~\ref{fig:profile_m_w}, when the price trajectory displays a pronounced increase, the function $F_1(t_c,m,\omega)$ almost always 
presents just one minimum along the $m$ direction and 3-4 local minima along the $\omega$ direction in the range $2<\omega<20$
(which may actually be relevant to capture higher harmonics of the logperiodicity structure \citep{ZhouNonpara03,ZhouSorRen03}).
Further, this method allows one to avoid sloppy directions in the $(t_c,\omega)$ plane, where the cost-function has a very long valley along the diagonal $t_c\sim \omega$, as illustrated in Figure~3b of~\citep{FilimonovSornette2011_LPPL_calibration}.

At the expense of a small increase of computational complexity, 
beyond its simplification, the cost-function given by equation~\eqref{eq:S2_new} provides
a substantial improvement in inference from the model. Namely, in addition to the point estimate~\eqref{eq:S_arg2_new}, 
expression~\eqref{eq:S2_new} allows one to analyze the whole profile of the cost function $F_2(t_c)$ and 
the dependence of the estimates $\hat m$ and $\hat\omega$ as a function of the critical time~$t_c$. In particular, one can identify 
all the extrema of $F_2(t_c)$ and their corresponding $m(t_c)$ and $\omega(t_c)$, 
from which expert judgment of the plausible scenarios can follow.

\begin{figure}[p!]
 \vspace{-0.8cm}
  \centering
  \includegraphics[width=0.8\textwidth]{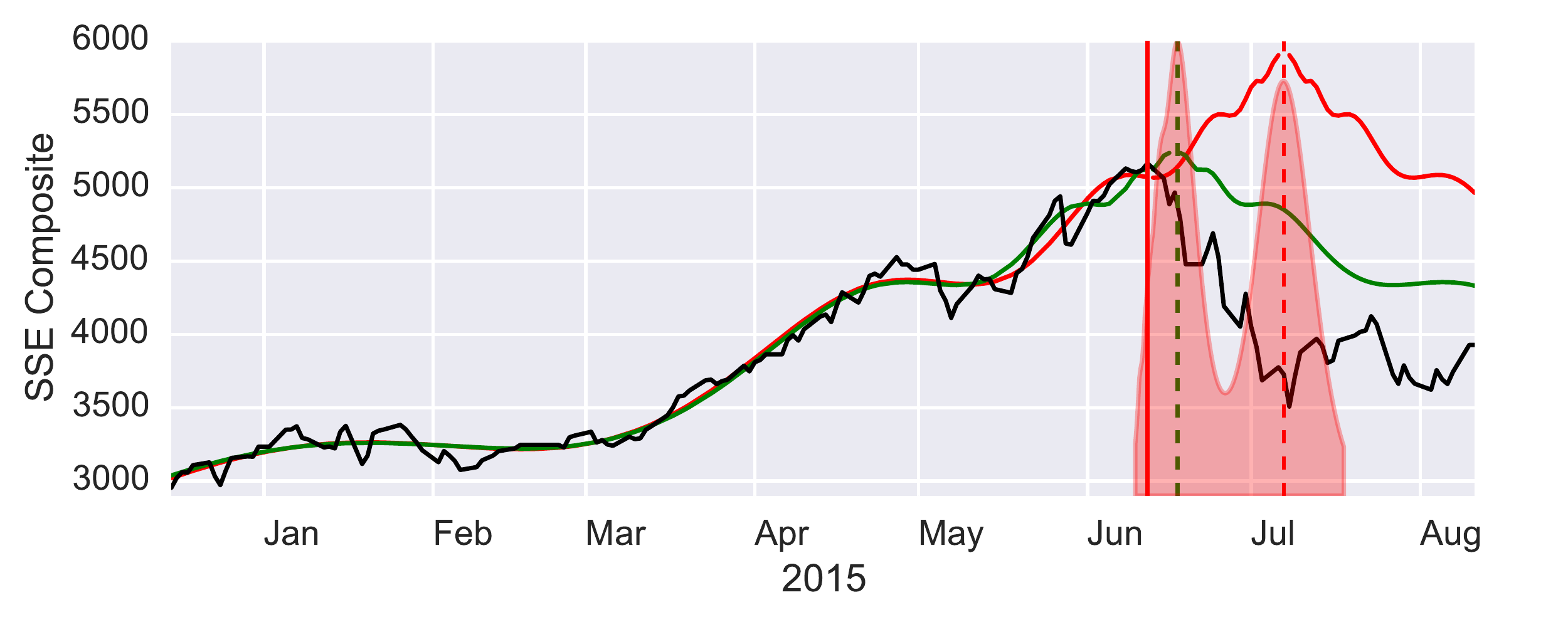}
  \caption{Price trajectory of the SSE Composite Index during the bubble of 2014--2015. The red vertical line denotes the date of 
  the analysis ($t_2=$ 2015-06-12).  Red and green solid lines correspond to the best and second best (see Figure~\ref{fig:sse_cost}) LPPLS fit in the window [2014-12-15, 2015-06-12] and their extrapolations to $t>t_2$. The vertical red and green dashed lines 
indicate the position of the critical times $t_c$ for these two fits:  2015-07-08 and 2015-06-18 respectively. The shaded red areas 
delineate the likelihood interval of $t_c$ at a 5\% cutoff together with the shape of the modified profile likelihood (see Sections~\ref{sec:modified}--\ref{sec:confidence} and Figure~\ref{fig:modified_profile}).} 
  \label{fig:price}
  \includegraphics[width=0.9\textwidth]{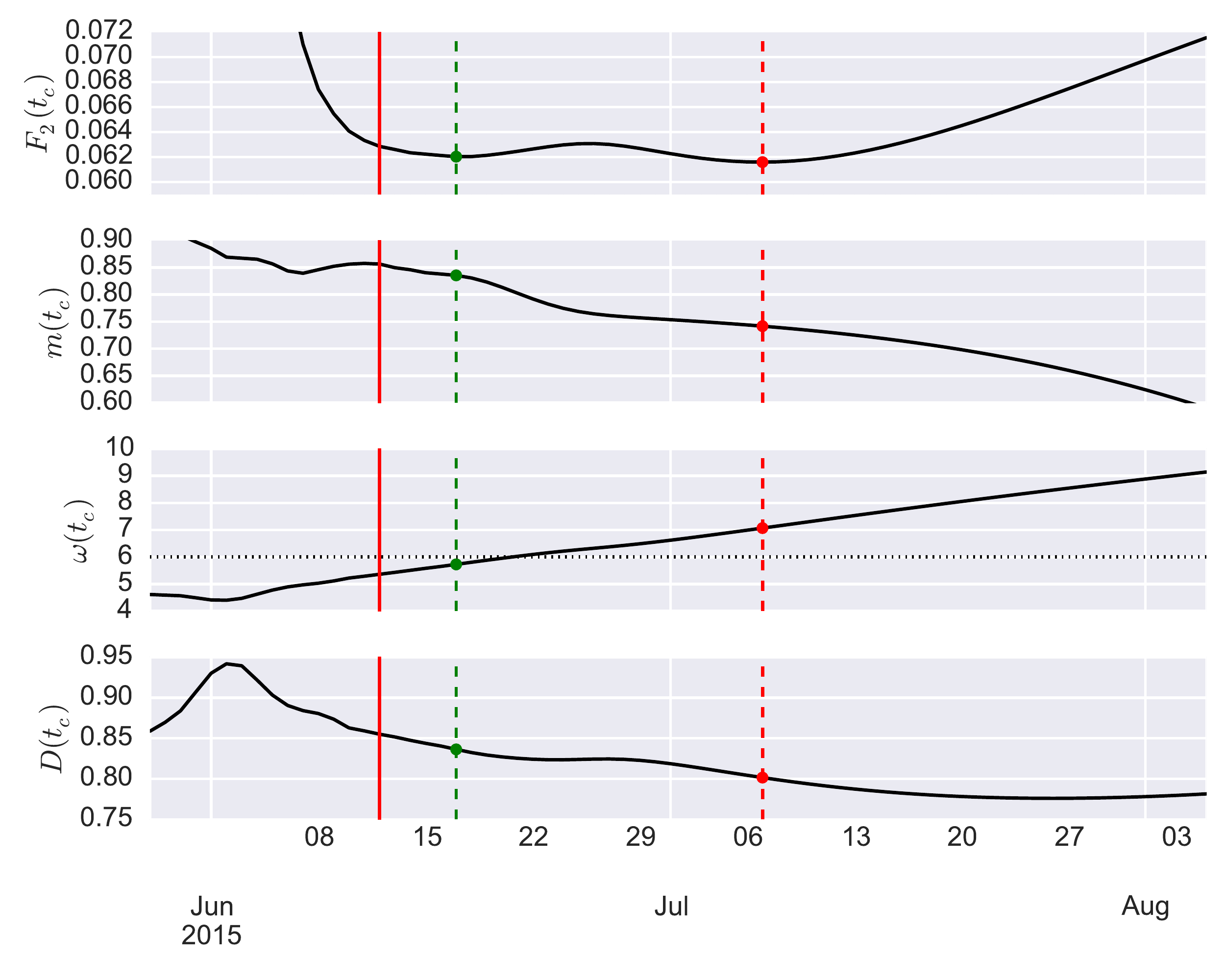}
  \caption{Profiles of the cost function $F_2(t_c)$ and parameters $\hat m$, $\hat \omega$ and damping $\hat D={m|B|}/{\omega|C|}$
as a function of $t_c$. The solid red vertical line indicates the date of analysis ($t_2=$ 2015-06-12), the red and green dotted vertical lines correspond to the dates of the best and alternative solutions (2015-07-08 and 2015-06-18 respectively). The horizontal dotted line 
gives the level of the threshold~\eqref{eq:constrains} for the logperiodic angular frequency parameter ($\omega>6$) that separates ``qualified'' fits from ``not qualified'' ones. The filled red and green circles show the point estimates of the model parameters --- for the best and alternative fits, respectively.} 
  \label{fig:sse_cost}
\end{figure}

As an example, we consider the recent bubble and following collapse of the Chinese market, when 
the Shanghai Composite Index (SSE Composite) appreciated by approximately 150\% between mid-2014 and mid-2015, 
peaked on June 12, 2015 and then lost 32\% to its first well-defined bottom reached on July 8, 2015. 
This bubble was detected by the Financial Crisis Observatory (FCO) at ETH Z\"urich and further documented and dissected in~\citep{Sornette_etal2015_Shanghai}. We use data provided by Thomson Reuters Dataworks Enterprise (see Section~\ref{sec:casestudies} for discussions). Figure~\ref{fig:price} presents the dynamics of the SSE Composite index 
together with the best LPPLS fit according to the OLS regression within the time window of $t_2-t_1=180$ calendar days ending at the date of $t_2=$June 12, 2015 when the market peaked.

In order to understand the ``microstructure'' of LPPLS fits, we employ the three-step subordination procedure
~\eqref{eq:S1_new}--\eqref{eq:yfgh}, \eqref{eq:S_arg2_new}--\eqref{eq:S2_new} and study the dependence of the cost-function $F_2(t_c)$ as well as $\hat m(t_c)$, $\hat \omega(t_c)$ and damping $\hat D(t_c)$ (see Fig.~\ref{fig:sse_cost}).
One can see that the global (best) solution with estimated critical time $\hat t_c$ of July 7, 2015 (with $F_2=0.0597$) is not the only minimum, and a second local minimum is found at $t_c =$ June 18, 2015 (with $F_2=0.0604$), which suggests a second plausible scenario. Despite almost identical values of the cost functions (sum of squared errors of residuals), we might reject the suboptimal solution 
on the basis of the fact that its logperiodic angular frequency falls outside of the empirical constraint~\eqref{eq:constrains} ($\hat\omega=7.18$ for the optimal solution and $\hat\omega=5.85$ for the suboptimal). Both solutions are associated with 
damping parameters that are below the constraint $D \geq1$ ($\hat D=0.8$ for the optimal and $\hat D=0.83$ for the suboptimal
solution), but they are both compatible with the relaxed constraint~\eqref{eq:constrains} (note that the value for the optimal solution 
is very close to the boundary of this constraint).


This case study exemplifies the essence of the problem of dealing with multiple and almost equivalent optimal solutions that
point to quite different future scenarios. Above we have invoked previous experience 
\citep{SornetteJohansen2010} to reject the second scenario.
However, this is not fully satisfactory from a theoretical view point. Moreover, past experience can be tainted by the 
use of the sub-optimal calibration procedure based on the original formulation of the model~\eqref{eq:lppl}.
To boot, past experience may not contain all possible situations, and surprises that are superficially of the 
``unknown unknown'' type \citep{Knight21,Taleb2007} from the point of view 
of past experience might actually be understandable and 
knowable with the appropriate conceptual and theoretical framework \citep{Sornette2009}.

The question we further investigate below is: How can we resolve between these two scenarios 
if we do not have (or do not want or trust to use) any prior information on what are plausible parameter values? 
In other words, how can we provide a quantitative estimation of how much one scenario is less likely than another? 
 
\section{Likelihood and Profile Likelihood \label{sec:likelihood}}

The OLS regression~\eqref{eq:OLS} represents the so-called \emph{normal estimation} of the model parameters, i.e.
provides \emph{Maximum Likelihood Estimates (MLE)} under the assumption that the error term $\varepsilon(\tau_i;t_c,\psi)$
is normally distributed. The likelihood has then a well-known form:
\begin{equation}\label{eq:lik}
	L(t_c,\psi,s)=(2\pi s)^{-n/2}\exp\left(-\frac{\sse(t_c,\psi)}{2s}\right)\rightarrow \max_{t_c,\psi,s},
\end{equation}
where $s=\sigma^2$ is a variance of the residuals $\varepsilon(\tau_i;t_c,\psi)$ and $n$ is the number of data points. 
By definition the likelihood is meaningful only up to an arbitrary positive constant, 
thus below we will omit such constant pre-factors. The MLE of the parameters $\{\hat t_c,\hat\psi\}$ is obtained straight from~\eqref{eq:lik}: considering the logarithm of the likelihood ($\ln L(t_c,\psi,s)$), one immediately arrives at~\eqref{eq:OLS} and an estimate for $\sigma^2$ is
\begin{equation}\label{eq:est_s}
	\hat\sigma^2\equiv\hat s=\frac1n \sse(\hat t_c, \hat \psi).
\end{equation}

Despite the equivalence of the MLE and OLS approaches in terms of computations, the MLE requires an explicit distributional assumption 
for the error term $\varepsilon$. This implies that the inference of $\psi$ is implicit with the likelihood approach, while further 
work with some sampling method is needed in the least squares approach.

As discussed above, we are mostly interested in the inference of the critical time $t_c$ while 
the other parameters $\eta=\{\psi,s\}\equiv\{m,\omega,A,B,C_1,C_2,s\}$ can be considered as \emph{nuisanse} 
parameters that are useful insofar that they allow to adapt the model to the variability of the data. 
The elimination of nuisance parameters is a well-known statistical problem, which
amounts to concentrating the likelihood around a single parameter of interest while accounting
for the extra uncertainty resulting from the estimation of the nuisance parameters.
Unfortunately, there is no technique that is efficient for all situations~\citep{BayarriDeGroot1992}, in particular because it is not always meaningful to discuss the uncertainty in one parameter independently from that of all others.

As already mentioned, in the Bayesian approach, the elimination of the nuisance parameters 
corresponds to integrating them out. However, the likelihood is not a regular density function and does not obey probability laws.
Therefore, the naive integration of the likelihood is not a meaningful operation. The proper implementation of the Bayesian 
approach requires specifying the prior distribution of all parameters $\{t_c,\eta\}$, calculating the posterior and then integrating out the nuisance parameters $\eta$ from the posterior to derive the posterior marginal distribution of $t_c$. The major limitation of the Bayesian approach is indeed a specification of the prior. We will not pursue this way directly, however, as shown in Section~\ref{sec:modified}, we will be able to capture the idea of integration over the nuisance parameters within a non-Bayesian framework.

One commonly used practice of elimination of nuisance parameters is based on a factorization of the complete likelihood into a product of the so-called \emph{marginal} and \emph{conditional likelihood functions}~\citep{KalbfleischSprott1970}. When available, this approach results in a genuine likelihood, i.e. the genuine probability of the observed data conditional on the parameter of interest ($t_c$). However, this approach requires transforming the sufficient statistics into a minimal sufficient statistics that has to be factored into two terms $T$ and $A$.
One of these terms, either the marginal distribution of $T$ or the conditional distribution of $T$ conditioned on $A$ (which is then called \emph{ancillary} for $t_c$), depends only on $t_c$, but not on $\eta$ (see discussions in~\citep{Pawitan2001} and for example~\citep{Basu1977,Severini1998a,Qin2005}). Given that the LPPLS model~\eqref{eq:lppl2} is highly nonlinear, it 
is not possible to find such factorization.

A simpler method is to construct the so-called \emph{profile likelihood}, which consists in replacing the nuisance parameters by their MLE at each fixed value of the parameter of interest. Given the joint likelihood~$L(t_c,\eta)$, the profile likelihood $L_p(t_c)$ is defined as
\begin{equation}\label{eq:profile}
	L_p(t_c)=\max_\eta L(t_c,\eta)\equiv L(t_c,\hat\eta_{t_c}),
\end{equation}
where $\hat\eta_{t_c}=\arg\max_\eta L(t_c,\eta)$ is a MLE for $\eta$ for a fixed value of $t_c$. The profile likelihood is often treated as a regular likelihood for further inference of $t_c$, i.e. one can normalize it, compute likelihood intervals or compare likelihood ratios.

The profile likelihood approach is technically identical to the analysis of the profile cost function $F_2(t_c)$ discussed in Section~\ref{sec:ols} Indeed, the MLE of $\hat\psi_{t_c}$ is given by the solution of the OLS~\eqref{eq:OLS}: the estimates of $\hat m_{t_c}$ and $\hat \omega_{t_c}$ are derived from~\eqref{eq:S2_new} where $\hat A_{t_c},\hat B_{t_c},\hat C_{1,t_c},\hat C_{2,t_c}$ are given by~\eqref{eq:ABC12}. Finally, the form of $\hat s_{t_c}$ is similar to~\eqref{eq:est_s}, where $\hat t_c$ is replaced by $t_c$. Moreover, the value of $L_p(t_c)$ can be directly derived from $F_2(t_c)$. Indeed, the estimation of $\hat s_{t_c}$ can be represented as
\begin{equation}\label{eq:est_s_tc}
	\hat s_{t_c}=\frac 1n\sse(t_c,\hat\psi)\equiv\frac1n F_2(t_c),,
\end{equation}
and, after plugging~\eqref{eq:est_s_tc} to~\eqref{eq:lik} according to~\eqref{eq:profile} we obtain:
\begin{equation}\label{eq:logLp}
	L_p(t_c)\propto \big(\hat s_{t_c}\big)^{-n/2}
	\propto \Big(F_2(t_c)\Big)^{-n/2},
\end{equation}
where we have omitted all constant terms. 

\begin{figure}[t!]
  \centering
  \includegraphics[width=0.9\textwidth]{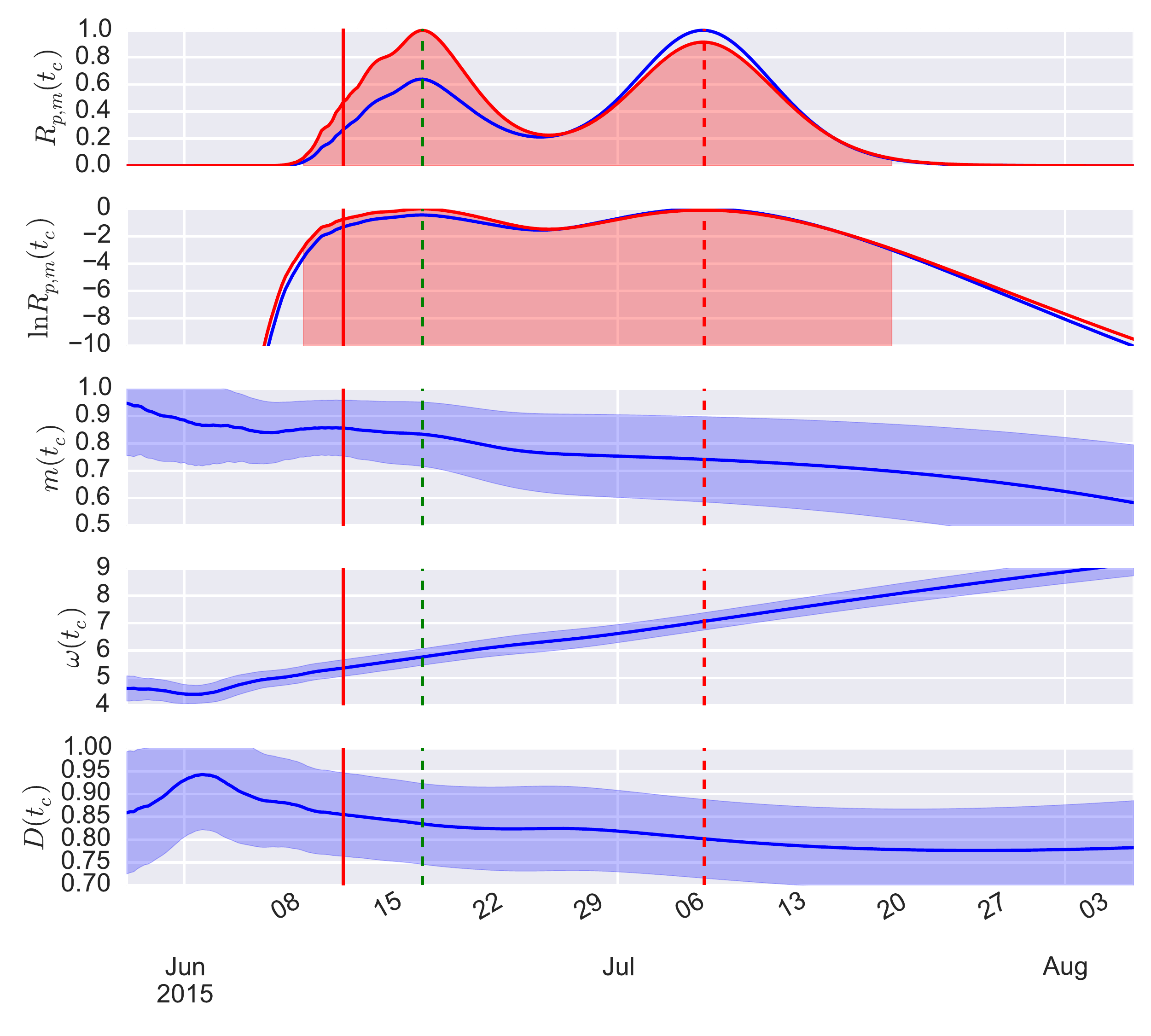}
  \caption{The top panel shows the relative profile likelihood (blue) and modified profile likelihood (red) as
a function of the critical time $t_c$. The second panel from the top shows the corresponding log-likelihoods. 
The red shaded area corresponds to the likelihood interval of $t_c$ at the 5\% probability level (see Section~\ref{sec:confidence}). 
The three bottom panels give the point MLE parameter estimates of the model, $m$, $\omega$ and $D$, as a function of the critical time $t_c$ (same as in Figure~\ref{fig:sse_cost}). Blue lines present point MLE estimates of the parameters and blue shaded areas correspond to their approximated likelihood intervals at the 5\% probability level (see Section~\ref{sec:m_w}).  
 The red continuous vertical line indicates the date of analysis ($t_2=$ 2015-06-12).
 The dashed red and green vertical lines correspond to the dates of the best and alternative solutions (2015-07-08 and 2015-06-18 respectively).} 
  \label{fig:modified_profile}
\end{figure}

Since the likelihood~\eqref{eq:lik} is meaningful only up to a constant, one usually considers the \emph{relative likelihood} (respectively, \emph{relative profile likelihood} or \emph{relative modified profile likelihood} that will be defined later), which 
is normalized to $1$ by its maximum and thus takes value in $[0,1]$:
\begin{equation}\label{eq:rel_lik}
	R(t_c)=\frac{L(t_c)}{\max_{t_c}L(t_c)}.
\end{equation}

Figure~\ref{fig:modified_profile} (blue curves) presents an example of the relative profile likelihood $R(t_c)$ calculated for the case discussed in Section~\ref{sec:ols} and presented in Figure~\ref{fig:sse_cost}. One can observe the same two extrema found  
with function $F_2(t_c)$, which correspond to very close values of the likelihood ($R_p(t_c^{(1)})=1$ for the best solution $t_c^{(1)}=$ 2015-07-08 and $R_p(t_c^{(2)})=0.64$ for the alternative solution $t_c^{(2)}=$ 2015-06-18). The likelihood ratio $R_p(t_c^{(1)})/R_p(t_c^{(2)})=1.56$ is not large enough to warrant preferring one maximum over the other.
The inference based on a point OLS (or MLE) estimate can thus be quite misleading. In fact, the interval of ``acceptable'' values for $t_c$ (the likelihood interval to be discussed in Section~\ref{sec:confidence}) is very broad, which confirms that a point estimation 
is far from reflecting the full picture.

\section{Modified Profile Likelihood \label{sec:modified}}

\subsection{General form of the modified profile likelihood}

As discussed above, the profile likelihood is often treated as a regular likelihood but in fact it is not a genuine likelihood function. Specifically, it treats the nuisance parameters at a fixed value $\hat \eta_{t_c}$ as if they were known. 
It may thus overstate the amount of information about $t_c$ and the inference on $t_c$ based on $L_p(t_c)$ may be grossly misleading if the data contain insufficient information about $\eta$ (in particular when $\eta$ is high-dimensional as in our case, which can lead to 
an \emph{overprecise} profile likelihood). Moreover, under certain conditions, the profile likelihood can provide unstable estimates with respect to small changes in the observed data. At the same time, more robust marginal and conditional likelihoods are not available 
in cases like ours.

In order to overcome this fundamental limitation of the profile likelihood, a series of adjusted versions have been proposed (see for instance, \citep{CoxReid1987,FraserReid1989,BarndorffNielsenCox1994,DiCiccio1996} for general discussions). Most of them require 
orthogonality between the parameter of interest ($t_c$) and nuisance parameter ($\eta$). In our case, orthogonality does not hold and, in order to come up with a parametrization $\tilde \eta$ that would be orthogonal to $t_c$, one needs to solve a system of differential equations~\citep{CoxReid1987}, which is nearly impossible to do analytically in our multi-dimensional non-linear case. Then, the most flexible approach is arguably the one proposed by~\citet{BarndorffNielsen1983}, who introduced the so-called \emph{modified profile likelihood} as a higher-order approximation to either a marginal or a conditional likelihood function (both derivations are possible). 

The modified profile likelihood amounts to introducing an extra modulating factor $M(t_c)$ to the profile likelihood:
\begin{equation}\label{eq:modified_lik}
	L_m(t_c)=M(t_c)L_p(t_c)=
	\left|I(\hat\eta_{t_c})\right|^{-1/2}
	\left|\frac{\partial \hat\eta}{\partial\hat\eta_{t_c}}\right|L_p(t_c),
\end{equation}
where $\hat\eta_{t_c}$ is a MLE of the nuisance parameters $\eta$ at a fixed value of $t_c$; $I(\hat\eta_{t_c})$ is the corresponding \emph{observed Fisher information} matrix on $\eta$ assuming that $t_c$ is known:
\begin{equation}\label{eq:observed_I}
	I(\hat\eta_{t_c})=-\left.\frac{\partial^2\ln L(t_c,\eta)}{\partial\eta\partial\eta^T}
	\right|_{\eta=\hat\eta_{t_c}},
\end{equation}
where $\eta^T$ stands for the transpose of $\eta$;  $\partial \hat\eta/\partial\hat\eta_{t_c}$ denotes a matrix of the first partial derivatives of 
the full MLE of the nuisance parameters $\eta$ with respect to the MLE calculated at a fixed value of $t_c$; finally, $|\cdot|$ denotes the absolute value of a matrix determinant. Here and in the following, we assume that the parameters form a column vector, thus second order derivatives of the form \eqref{eq:observed_I} define a matrix.

The term $\left|I(\hat\eta_{t_c})\right|^{-1/2}$, which describes the curvature of the likelihood, can be considered as a penalty that subtracts from the profile log-likelihood ``undeserved'' information due to the estimation of the nuisance parameter $\eta$. And the Jacobian term $J(t_c)=|\partial \hat\eta/\partial\hat\eta_{t_c}|$ is needed to make the modified profile likelihood invariant with respect to the transformations of the nuisance parameters~\citep{Pawitan2001}. Practically, this term is extremely difficult to evaluate, which dramatically limits the application of~\eqref{eq:modified_lik}.

Unlike profile likelihoods, the modified profile likelihood is a genuine likelihood function and has a number of important properties. First, as just mentioned, due to the Jacobian term, $L_m(t_c)$ is invariant with respect to a reparametrisation of the model (such as the variable change~\eqref{eq:C12}). Second, it does not require orthogonality of $t_c$ and $\eta$, neither does it require specification of an ancillary statistics. Finally, \citet{Severini2007} has shown that the modified profile likelihood can be considered as an approximation to a class of integrated likelihood functions and very naturally arises from a non-Bayesian inference with integrated likelihood. But, in contrast to the Bayesian approach or integrated likelihood functions, the modified profile likelihood does not require specification of a prior density for the nuisance parameters --- the main limitation that hampered us from pursuing this direction. 

\subsection{Inference on the errors variance}\label{ssec:sigma}

The importance of the modified profile likelihood cannot be overstated, given that it is considered one of the breakthroughs in modern parametric inference~\citep{DiCiccio1997}. Perhaps the best illustration of the power of this method relates to the estimation of the variance $s=\sigma^2$ in nonlinear regressions such as~\eqref{eq:lik}. It is well known that the standard estimation~\eqref{eq:est_s} or~\eqref{eq:est_s_tc} is biased and it should be corrected to account for the number of degrees of freedom, i.e. the number of free parameters to estimate. The modified profile likelihood provides this correction as follows.

For the time being, let us consider $s$ as a parameter of interest and all the other parameters $\lambda=\{t_c,m,\omega,A,B,C_1,C_2\}$ as nuisance parameters. Parameters $s$ and $\lambda$ are not only informationally orthogonal, but the estimation of $\lambda$ does not depend on $s$ at all (since $\lambda$ is given straightforwardly from the OLS method). Thus, $\hat\lambda_s\equiv\hat\lambda$ and $|\partial\hat\lambda/\partial\hat\lambda_s| \equiv1$. Having taken care of the Jacobian, we only need to calculate the observed Fisher information~$I(\hat s_\lambda)$. 

Straight from~\eqref{eq:lik}, we can derive the vector of first derivatives of the log-likelihood --- the so-called \emph{score function} $S(\lambda)$:
\begin{equation}\label{eq:dL_lambda}
	S(\lambda)=\frac{\partial \ln L(s,\lambda)}{\partial\lambda}=
	-\frac1{2s}\frac{\partial \sse(\lambda)}{\partial\lambda}.
\end{equation}
The negative second derivative gives us the observed Fisher information matrix whose determinant reads
\begin{equation}\label{eq:d2L_lambda}
	|I(\lambda)|=\left|-\frac{\partial^2 \ln L(s,\lambda)}{\partial\lambda\partial\lambda^T}\right|=
	\left(\frac1{2s}\right)^{p_\lambda}\left|\frac{\partial^2 \sse(\lambda)}{\partial\lambda\partial\lambda^T}\right|,
\end{equation}
where $p_\lambda=\dim\lambda=7$ is the dimension of the nuisance parameter space. Before plugging the expression~\eqref{eq:d2L_lambda} into~\eqref{eq:modified_lik} in order to obtain the modified profile likelihood of $s$, notice that (i) the matrix of second-order derivatives $\partial^2 \sse(\lambda)/\partial\lambda\partial\lambda^T$ in~\eqref{eq:d2L_lambda} does not depend on the parameter of interest $s$ explicitly and (ii) the OLS estimation $\hat\lambda_s\equiv\hat\lambda=\{\hat t_c,\hat m,\hat \omega,\hat A,\hat B,\hat C_1,\hat C_2\}$ also does not depend on $s$. Thus, the determinant of $\partial^2 \sse(\lambda)/\partial\lambda\partial\lambda^T$
is a constant with respect to the variable $s$ and therefore can be omitted. Then, 
the modified profile likelihood of $s$ can be expressed in the following form:
\begin{equation}\label{eq:Lm_s}
	L_m(s)\propto s^{(n-p_\lambda)/2}\exp\left(-\frac{\sse(\hat\lambda)}{2s}\right),
\end{equation}
which leads to the following MLE for $s$:
\begin{equation}\label{eq:est_s_unbiased}
	\hat s=\frac1{n-p_\lambda} \sse(\hat \lambda).
\end{equation}
The denominator $n-p_\lambda$, which is different from $n$ in~\eqref{eq:est_s}, not only removes the bias of the estimator, 
but also results in a better likelihood-based inference of $s$ when it is needed.

\subsection{Approximation of the modified profile likelihood}\label{ssec:approx_modified}

Given all the remarkable properties of the modified profile likelihood, it has one very serious limitation, briefly mentioned above. Namely, for many realistic models, it is extremely difficult to calculate the Jacobian in~\eqref{eq:modified_lik}.
In order to get an intuition about the nature of the difficulty, it is useful to express it in the following form (see e.g.~\citep{Pawitan2001}):
\begin{equation}\label{eq:Jacobian}
	J(t_c)\equiv\left|\frac{\partial \hat\eta}{\partial\hat\eta_{t_c}}\right| =
	\frac{|I(\hat\eta_{t_c})|}{|C(t_c,\hat\eta_{t_c};\hat t_c, \hat\eta)|},
\end{equation}
where the matrix $C(t_c,\hat\eta_{t_c};\hat t_c, \hat\eta)$ is given by the second-order derivatives of a log-likelihood
$L(t_c,\hat\eta_{t_c};\hat t_c,\hat\eta,a)$ that includes a new parameter 
$a$ that is ancillary for $\{\hat t_c,\hat\eta\}$, i.e. $\{\hat t_c,\hat\eta,a\}$ is a sufficient statistic of the model:
\begin{equation}\label{eq:C}
	C(t_c,\hat\eta_{t_c};\hat t_c, \hat\eta)=
	\frac{\partial^2 \ln L(t_c,\hat\eta_{t_c};\hat t_c,\hat\eta,a)}
	{\partial\hat\eta_{t_c}\partial\hat\eta^T}.
\end{equation}
In contrast to the observed Fisher information, which is also defined as a second-order derivative~\eqref{eq:observed_I} calculated 
at a specific MLE $\hat\eta_{t_c}$, the calculation of $C$~\eqref{eq:C} is much more complicated because, in the general case, it requires a reformulation of the log-likelihood in order to introduce an explicit dependence on the MLEs $\hat\eta_{t_c}$ and $\hat\eta$. 
In the case of inference of the variance $s$ presented in Section~\ref{ssec:sigma}, we 
used the orthogonality of $s$ and $\lambda$, which resulted in $\hat\lambda_s\equiv\hat\lambda$.
In contrast, for the inference on $t_c$, there is no closed form expression for $J(t_c)$. 
And, as discussed above, we cannot use the \emph{adjusted profile likelihood}~\citep{CoxReid1987} because orthogonalization of the nuisance parameters with respect to $t_c$ is not feasible either.

In order to calculate expression \eqref{eq:modified_lik}, several approximation of $L_m$ were proposed (see e.g.~\citep{BarndorffNielsen1994,Skovgaard1996,Severini1998,Fraser1999,Skovgaard2001} and~\citep{Severini2000_book,PaceSalvan2006} for reviews).
We will use the approximation to the modified profile likelihood proposed by~\citet{Severini1998}. This approximation requires only the covariance of score functions of the nuisance parameters and is thus fairly easy to compute. As shown in~\citep{Severini1998}, this approximation is invariant under the reparametrization of the model, is stable in the sense of conditional inference and agrees with the exact $J(t_c)$~\eqref{eq:Jacobian} to order $\mathcal{O}(n^{-1})$ in the moderate deviation sense and to order  $\mathcal{O}(n^{-1/2})$ in the large deviation sense, where $n$ is the number of data points. Another famous approximation by~\citet{BarndorffNielsen1994} agrees with the exact form of $L_m$ only to $\mathcal{O}(1)$ in the large deviation sense and thus is not asymptotically better than the simple profile likelihood $L_p$.

\citet{Severini1998} suggested to approximate the matrix~\eqref{eq:C} with the  covariance matrix of score functions of the following form:
\begin{equation}\label{eq:covariance}
	C(t_c,\hat\eta_{t_c};\hat t_c, \hat\eta)\approx
	\Sigma(t_c,\hat\eta_{t_c};\hat t_c, \hat\eta)
\end{equation}
where
\begin{equation}\label{eq:covariance}
	\Sigma\left(t_{c;1},\eta_{1};t_{c;2},\eta_{2}\right)=\mathrm{E}_{(2)}\left[
	\left.\frac{\partial \ln L(t_c,\eta)}{\partial\eta}\right|_{\substack{t_c=t_{c;1}\\\eta=\eta_{1}}}
	\left.\frac{\partial \ln L(t_c,\eta)}{\partial\eta^T}\right|_{\substack{t_c=t_{c;2}\\\eta=\eta_{2}}}
	\right].
\end{equation}
Here the expectation $\mathrm{E}_{(2)}[\cdot]$ is taken with respect to the probability distribution of error term $\varepsilon(\tau; t_{c;2},\eta_{2})$ that corresponds to the parameters $\{t_{c;2},\eta_{2}\}$.
In contrast to the exact form \eqref{eq:C}, here we need only the score functions, which have expressions
similar to \eqref{eq:dL_lambda}. When calculation of~\eqref{eq:covariance} is too complicated, one can exploit the independence of observations $\varepsilon(\tau_i;t_c,\eta)$ and replace the covariance matrix~\eqref{eq:covariance} by its asymptotically equivalent sample estimation~\citep{Severini1999}:
\begin{equation}\label{eq:covariance_hat}
	\widehat\Sigma\left(t_{c;1},\eta_{1};t_{c;2},\eta_{2}\right)=
	\sum_{i=1}^n
	\left.\frac{\partial f(\tau_i; t_c, \eta)}{\partial\eta}\right|_{\substack{t_c=t_{c;1}\\\eta=\eta_{1}}}
	\left.\frac{\partial f(\tau_i; t_c, \eta)}{\partial\eta^T}\right|_{\substack{t_c=t_{c;2}\\\eta=\eta_{2}}},
\end{equation}
where
\begin{equation}\label{eq:f}
	f(\tau;t_c,\eta)=
	-\frac 12\ln(2\pi s)-\frac 1{2s}\big(\ln p(\tau)-\lppls(\tau;t_c,\psi)\big)^2
\end{equation}
is a contribution from an individual observation to the log-likelihood.
Of course, the adjustment~\eqref{eq:covariance} based on the theoretical covariance is superior to the sample-based estimation~\eqref{eq:covariance_hat}, in particular in cases of small sample \mbox{size~\citep{Severini1999,BesterHansen2009}}. For our purposes, we will use the exact form \eqref{eq:covariance}, which can be calculated in closed form. Finally, plugging~\eqref{eq:covariance} into~\eqref{eq:Jacobian} and~\eqref{eq:modified_lik}, we obtain the desired approximated expression for $L_m(t_c)$:
\begin{equation}\label{eq:modified_lik_approx}
	L_m(t_c)\approx
	\frac{\left|I(\hat\eta_{t_c})\right|^{1/2}}
	{\left| \Sigma(t_c,\hat\eta_{t_c};\hat t_c, \hat\eta)\right|} L_p(t_c)	~.
\end{equation}

In this expression (\ref{eq:modified_lik_approx}), the profile likelihood $L_p(t_c)$ is given 
by the previously calculated expression \eqref{eq:logLp}. The observed Fisher information $I(\hat\eta_{t_c})$
and the covariance matrix $ \Sigma(t_c,\hat\eta_{t_c};\hat t_c, \hat\eta)$ are given in Appendix~\ref{app:derive}. 
Omitting terms that do not depend on $t_c$, the final expression for $L_m(t_c)$ is given by:
\begin{equation}\label{eq:Lm_LPPLS}
	L_m(t_c)\propto 
	\frac{\displaystyle \big(\hat s_{t_c}\big)^{-(n-p-2)/2}
	\left| \sum_{i=1}^n\frac{\partial^2 \lppls(\tau_i;t_c,\psi)}{\partial \psi\partial \psi^T}\right|^{1/2}_{\psi=\hat\psi_{t_c}}
	}{\displaystyle
	\left| \sum_{i=1}^n 
		\left.\frac{\partial \lppls(\tau_i;t_c,\psi)}{\partial\psi}\right|_{\substack{t_c=t_{c}\\\psi=\hat\psi_{t_c}}}
		\left.\frac{\partial \lppls(\tau_i;t_c,\psi)}{\partial\psi^T}\right|_{\substack{t_c=\hat t_{c}\\\psi=\hat\psi}}
	\right|},
\end{equation}
where $p=\dim\psi=6$. Following~\citep{Severini1999}, let us introduce the rectangular $n\times p$ matrix
\begin{equation}\label{eq:X}
	X_{ij}(t_c,\psi)=\frac{\partial \lppls(\tau_i;t_c,\psi)}{\partial \psi_j}
\end{equation}
and the square $p\times p$ matrix
\begin{equation}\label{eq:H}
	H_{ij}(t_c,\psi)=\sum_{k=1}^n
	\big(\ln p(\tau_k)-\lppls(\tau_k;t_c,\psi)\big)
	\frac{\partial^2\lppls(\tau_k;t_c,\psi)}{\partial \psi_i\partial \psi_j},
\end{equation}
where $\psi_j$ denotes the $j$-th element of the nuisance parameter vector $\{m,\omega,A,B,C_1,C_2\}$. 
Then, expression~\eqref{eq:Lm_LPPLS} simplifies into
\begin{equation}\label{eq:Lm_LPPLS_matrix}
	L_m(t_c)\propto 
	\frac{	\big| X^T(t_c,\hat\psi_{t_c}) X(t_c,\hat\psi_{t_c}) - H(t_c,\hat\psi_{t_c}) \big|^{1/2}}
	{\big| X^T(\hat t_c,\hat\psi) X(t_c,\hat\psi_{t_c})\big|}\big(\hat s_{t_c}\big)^{-(n-p-2)/2},
\end{equation}
where $\hat s_{t_c}$ is the MLE estimate of the variance~\eqref{eq:est_s_tc} (it is not the adjusted estimate~\eqref{eq:est_s_unbiased}), $\hat \psi_{t_c}$ is a vector of MLE estimates for the LPPLS parameters at a fixed value of $t_c$ and $\{\hat t_c,\hat\psi\}$ are full MLE estimates of the parameters.
The expressions of the first-order and second-order partial derivatives that are needed for~\eqref{eq:X} and~\eqref{eq:H} are given by~\eqref{eq:LPPLS_partial_1} and~\eqref{eq:LPPLS_partial_2} in Appendix B.

As a concrete illustration, we consider the 2015 bubble in Chinese markets already discussed in Sections~\ref{sec:ols}--\ref{sec:likelihood}. The red curves in the two top panels of figure \ref{fig:modified_profile} show the modified profile likelihood 
obtained from expression (\ref{eq:Lm_LPPLS_matrix}). 
It is particularly interesting that the adjustments to $L_m(t_c)$ have significantly changed the picture, since the ``alternative'' extremum 
has now a higher likelihood than the best OLS solution ($R_m(t_c^{(2)})=1.$ versus $R_m(t_c^{(2)})=0.91$).
Thus, accepting the OLS point estimate would bias $\hat t_c$ by 19 days. 
The likelihood ratio is now even smaller than for the simple profile likelihood, $R_m(t_c^{(2)})/R_m(t_c^{(1)})=1.096$, and both extrema are almost equally likely.

\subsection{Likelihood Intervals and Confidence Intervals}\label{sec:confidence}

The major improvement of the standard MLE interpretation~\eqref{eq:lik} over the OLS~\eqref{eq:OLS} is the fact that MLE provides a direct estimation of the uncertainty in estimated parameters. In other words, MLE can provide not just the \emph{point} estimate of $\theta$ but a \emph{range estimate} of values that are possible given the observed data. Such inference is based on the \emph{likelihood ratio} $R(\theta)$,
introduced earlier in the form of the relative likelihood~\eqref{eq:rel_lik} defined as the ratio of the likelihood normalized by its maximum value. When $R(\theta_0)$ is sufficiently small, the hypothesis that the parameter could have a value $\theta=\theta_0$ can be rejected as ``unsupported by the data''.

However, the question of ``how small is sufficiently small?'' often does not have a rigorous solution and strongly depends on the problem. Many authors suggest to choose some rather arbitrary \emph{cutoff} and consider values of likelihood ratio above this cutoff to define a so-called \emph{likelihood-based confidence interval} or \emph{likelihood interval (LI)}. For example, many authors including~\citet{Fisher1956}suggested that parameter values $\theta$ for which $L(\hat\theta)/L(\theta)=1/R(\theta)>15$ should be declared ``implausible'', where $\hat\theta$ is the standard MLE. 

In regular one-parameter models, one can create a frequentist confidence interval, based on a probability-based calibration. For example, the log-likelihood ratio test statistic $-2\ln R(\theta)$ can be then approximated using Wilk's theorem, and an approximate p-value is given by the $\chi^2$-distribution with one degree of freedom. Further, for regular likelihood functions, i.e. those that are well-approximated by a quadratic function, one can define a \emph{confidence interval (CI)} around MLE $\hat\theta$ solely based on the observed Fisher information. For example, a standard error would have the form $I^{-1/2}(\hat\theta)$ and 95\% CI would be given by $\hat\theta\pm1.96I^{-1/2}(\hat\theta)$ (\emph{Wald confidence interval}).

For our applications, these approaches are not perfectly suited. First, as can be seen in Figure~\ref{fig:modified_profile}, the profile and modified profile likelihoods are not regular: they are asymmetric and can be even multi-modal, so that the Wald CI does not provide a meaningful representation of parameter uncertainty. For the same reason, the calibration of the distribution of the test statistic under the null hypothesis is not straightforward and would be computationally very difficult given the dimensionality of the parameter space and the complexity of the LPPLS model~\eqref{eq:lppl2}. Finally, within our domain of application, an interpretation of the frequentist probability-based confidence intervals is not very intuitive. Indeed, giving the idiosyncratic nature of a bubble, in order to make sense out of the probabilistic intervals, one needs to involve a many-worlds interpretation, where price trajectory is shared among multiple universes. 

For all the reasons mentioned above, we choose to operate with likelihood intervals that are more intuitive in our context and are not subjected to the assumptions of regularity. Following Fisher's suggestion, we define the likelihood interval at the 5\% cutoff:
\begin{equation}\label{eq:LI_tc}
	\mathrm{LI}(t_c)=\left\{t_c: R_m(t_c)=\frac{L_m(t_c)}{L_m(\hat t_c)}>0.05\right\}.
\end{equation}
The two top panels of Figure~\ref{fig:modified_profile} show such 5\% modified profile likelihood intervals for the case of 2015 Chinese bubble.

\section{Filtering and likelihood intervals for nuisance parameters \label{sec:m_w}}

Similarly to the inference on the critical time $t_c$, let us apply the modified profile likelihood approach 
to estimate the likelihood intervals (LIs) of parameters $m$ and $\omega$.
This is of interest in particular because $m$ and $\omega$ are used in 
the filtering conditions~\eqref{eq:constrains}. 

Three different ways of inference on $m$ and $\omega$ exist.
First, we could consider $m$ (respectively, $\omega$) as the sole parameter of interest and $\eta_m=\{t_c,\omega,A,B,C_1,C_2,s\}$ (respectively, $\eta_\omega=\{t_c,m,A,B,C_1,C_2,s\}$) as the vector of nuisance parameters. Then, an analysis similar to
that developed in Sections~\ref{sec:modified}--\ref{sec:confidence} would provide the corresponding modified profile likelihood and LIs for these two parameters. However, keeping in mind that the parameter of main interest is the critical time $t_c$, we would need to somehow associate the inferred LIs for $m$ ($\omega$) with the corresponding values of $t_c$. 

A second approach consists in targeting
the vector $\theta=\{t_c,m,\omega\}$, while $\eta_\theta=\{A,B,C_1,C_2,s\}$ becomes the vector of nuisance parameters. The general framework remains the same as before. However, the computational complexity increases substantially, since the modified profile likelihood $L_m(\theta)$ is a 3-dimensional function. And the analysis of such function is not straightforward, with many 
2D-cross-sections needed to obtain a suitable understanding of the topology in four dimensional space. Or we would need
another layer of profile or modified profile likelihood to be calculated.

Here, we employ a third approach. For any fixed value of $t_c$, we consider a reduced LPPLS formula that is parameterized solely with the vector $\{m,\omega,A,B,C_1,C_2\}$. We then calculate a modified profile likelihood $L_m(m;t_c)$ (respectively $L_\omega(\omega;t_c)$) with $\eta_m=\{\omega,A,B,C_1,C_2,s\}$ (respectively, $\eta_\omega=\{m,A,B,C_1,C_2,s\}$) as the vector of nuisance parameters. The expression for $L_m(\cdot)$ is then similar to~\eqref{eq:Lm_LPPLS_matrix}. For example, $L_m(m;t_c)$ has the form
\begin{equation}\label{eq:Lm_LPPLS_matrix_m}
	L_m(m;t_c)\propto 
	\frac{	\big| X_m^T(t_c,m,\hat\phi_{t_c,m}) X_m(t_c,m,\hat\phi_{t_c,m}) - 
			H_m(t_c,m,\hat\phi_{t_c,m}) \big|^{1/2}}
	{\big| X_m^T(\hat t_c,\hat m,\hat\phi) X_m(t_c,m,\hat\phi_{t_c,m})\big|}
	\big(\hat s_{t_c,m}\big)^{-(n-p-2)/2},
\end{equation}
where $\phi=\{\omega,A,B,C_1,C_2\}$, $p=\dim\phi=5$, $\{\hat t_c,\hat m,\hat \phi\}$ is the full MLE of all parameters of the LPPLS model and $\hat \phi_{t_c,m}$ is the MLE of $\phi$ at fixed values of $\{t_c,m\}$. Finally $\hat s_{t_c,m}=\sse(t_c,m,\hat\phi_{t_c,m})/n$ and the matrice  $X_m$ is obtained from $X$~\eqref{eq:X} by removing the first column and the matrix $H_m$ is the principal submatrix of $H$~\eqref{eq:H}, obtained by removing its first row and first column. Targeting $\omega$, the expression for $L_m(\omega;t_c)$ is also given by\eqref{eq:Lm_LPPLS_matrix_m} up to a replacement of $m$ by $\omega$, where $X_\omega$ is obtained from $X$ by removing the second column, and $H_\omega$ is obtained from $H$ by removing the second column and second row.

\begin{figure}[t!]
  \centering
  \includegraphics[width=0.9\textwidth]{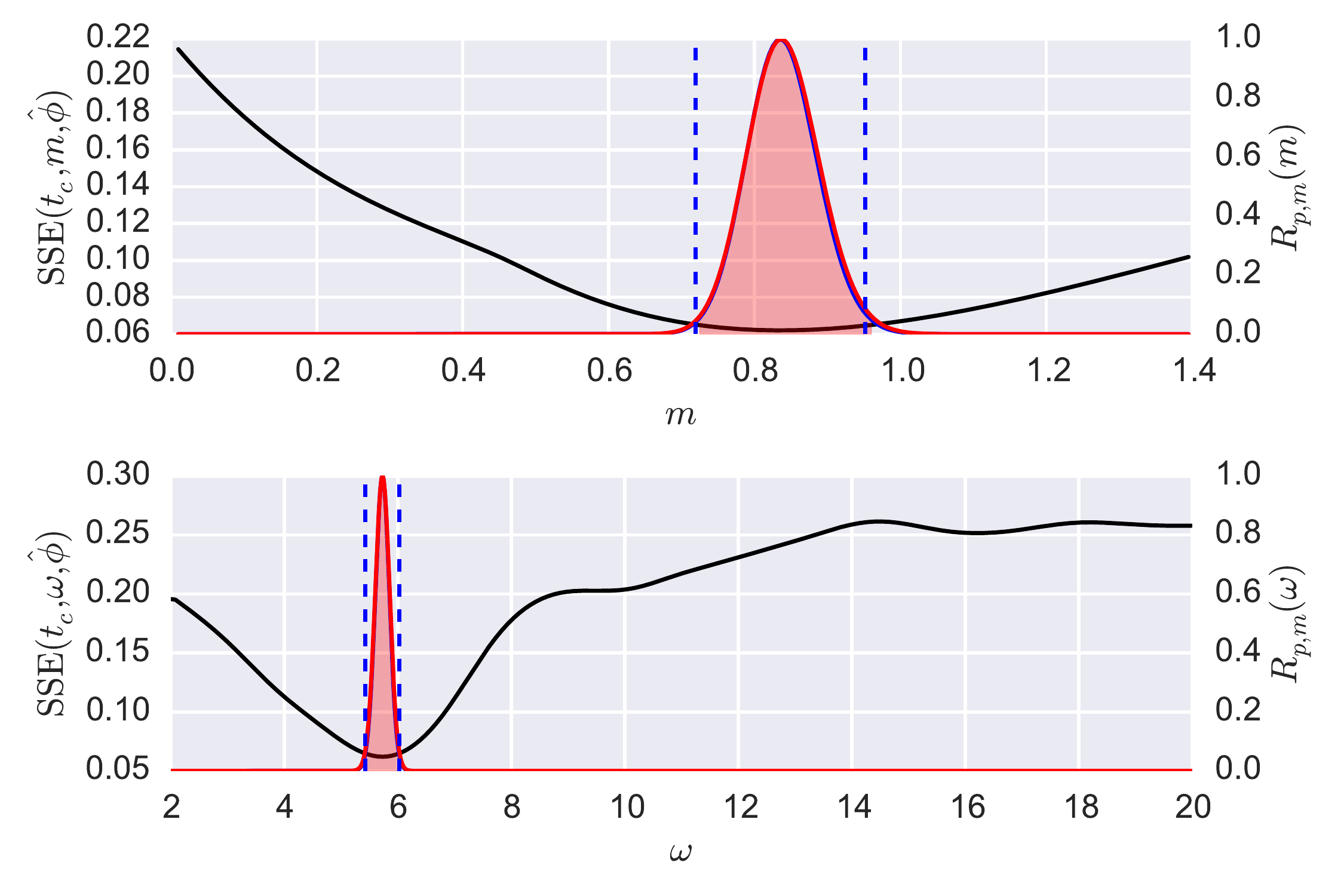}
  \caption{Profile of the cost function $F(\cdot)$ (black line, left scale), profile likelihood $L_p(\cdot)$ (blue line, right scale) and modified profile likelihood (red line, right scale) for the power law exponent $m$ (top panel) and logperiodic angular frequency $\omega$ (bottom panel) for $t_c=$2015-06-17. Note that the profile and modified profile likelihood almost coincide. The red shaded intervals represent the likelihood intervals $\mathrm{LI}(\cdot)$ at the 5\% cutoff. The vertical blue dashed lines delineate the approximated likelihood intervals~\eqref{eq:approx_LI_m_w} at the 5\% cutoff.} 
  \label{fig:profile_m_w}
\end{figure}

Figure~\eqref{fig:profile_m_w} presents the profile and modified profile likelihoods for the parameters $m$ and $\omega$ in the case considered before (Figures~\ref{fig:price}--\ref{fig:modified_profile}) for the fixed value of $t_c=$2015-06-17. It is interesting to note that the SSE profile of parameter $m$ at a fixed $t_c$ is unimodal in the range of interest. Moreover, our tests show that this is typically the case for a broad range of values $0<m\lesssim 3$. The SSE profile for $\omega$ is multimodal, but when the price trajectory exhibits a clear upward trend with a substantial price appreciation over the window of calibration $[t_1, t_2]$ (e.g. when the price increase is substantially larger than the volatility), then the best solution $\hat\omega$ is often clearly delineated and the likelihood profile is essentially unimodal, i.e. the alternative solutions are implausible (as in Figure~\ref{fig:profile_m_w}). 

In Figure~\eqref{fig:profile_m_w}, it is almost impossible to distinguish the profile likelihood 
from the modified profile likelihood in the visible range of values. The values at which 
the log-likelihoods start to disagree, i.e. for $R_m(\cdot;t_c)=\ln L_m(\cdot;t_c)/\ln L_m(\hat \cdot; t_c)\lesssim-15$, cannot be
been seen in this linear scale representation. We have found that this situation is typical for many 
other cases. This very close agreement means that the profile likelihood $L_p(\cdot;t_c)$ is already a good approximation to either 
the marginal or the conditional likelihood so that we could use it directly for the inference of likelihood intervals. Moreover, the peak of the profile likelihood can often be  well approximated by a quadratic function, allowing use to use this approximation for 
an analytical evaluation of LI\footnote{We need to mention that this is not always the case, and a bi-modal structure of both profiles on $m$ and $\omega$ is also possible, though rare. Moreover, in some cases, the second-order approximation of the modified profile likelihood might completely change the estimation of these parameters (see Appendix~\ref{sec:double_omega}).}. 

In contrast to the estimated likelihood, the negative curvature of the profile likelihood function of a parameter $\eta_i$ is not equal to $[I(\hat\eta_{t_c})]_{i,i}$, where $I$ is the observed Fisher information matrix~\eqref{eq:I_LPPLS_final}, but to $([I^{-1}(\hat\eta_{t_c})]_{i,i})^{-1}$ (see e.g. derivations in~\citep{HeldBove2013}). One can prove that $[I(\hat\eta_{t_c})]_{i,i}\ge([I^{-1}(\hat\eta_{t_c})]_{i,i})^{-1}$, which means that the observed Fisher information of the profile likelihood is smaller than or equal to the observed Fisher information on the estimated likelihood. This illustrates the fact that the nuisance parameter $\eta$ has to be estimated and thus adds to the uncertainty of the parameter of interest. Taking this approximation of the curvature into account, we can write the following Taylor expansion for the profile likelihood of $m$ and $\omega$ at a fixed $t_c$:
\begin{equation}\label{eq:approx_logL_m_w}
\begin{array}{rll}
	\ln L_p(m;t_c)&\approx\ln L(t_c,\hat\eta_{t_c})
	&-\frac12 ([I^{-1}(\hat\eta_{t_c})]_{1,1})^{-1}(m-\hat m_{t_c})^2,\\
	\ln L_p(\omega;t_c)&\approx\ln L(t_c,\hat\eta_{t_c})
	&-\frac12 ([I^{-1}(\hat\eta_{t_c})]_{2,2})^{-1}(\omega-\hat \omega_{t_c})^2,\\
\end{array}
\end{equation}
and thus the likelihood intervals at a cutoff of level $c$ are given by
\begin{equation}\label{eq:approx_LI_m_w}
\resizebox{.92\hsize}{!}{$
\begin{array}{rlll}
	\mathrm{LI}(m;t_c)&=\left\{m: \frac{L_p(m;t_c)}{L_p(\hat m; t_c)}>c\right\}
	&=\left\{m:|m-\hat m_{t_c}|<\Delta_{m;t_c}\right\},
	~&\Delta_{m;t_c}=\sqrt{-2\ln c\left[I^{-1}(\hat\eta_{t_c})\right]_{1,1}},\\
	\mathrm{LI}(\omega;t_c)&=\left\{\omega: \frac{L_p(\omega;t_c)}{L(\hat \omega;t_c)}>c\right\}
	&=\left\{\omega:|\omega-\hat \omega_{t_c}|<\Delta_{\omega;t_c}\right\},
	~&\Delta_{\omega;t_c}=\sqrt{-2\ln c\left[I^{-1}(\hat\eta_{t_c})\right]_{2,2}}.\\
\end{array}
$}
\end{equation}
Here, $I(\hat\eta_{t_c})$ has the form \eqref{eq:I_LPPLS_final} (Appendix A), and its submatrix of partial derivatives can be written in a matrix form similar to the numerator in~\eqref{eq:Lm_LPPLS_matrix}.
These likelihood intervals for $c=0.05$ are indicated with dashed vertical lines in Figure~\ref{fig:profile_m_w}, and one can see that they provide a very good approximation for the true LIs based on the modified profile likelihood for $m$ and $\omega$ at a fixed $t_c$ (red shaded areas).

The likelihood interval for the damping parameter $D=m|B|/\omega|C|$ is slightly more difficult to calculate. Because $D$ does not enter LPPLS expression~\eqref{eq:lppl2} directly, we first need to perform a variable change, e.g. by replacing the vector $\eta=\{m,\omega,A,B,C_1,C_2,s\}$ with $\zeta=\{D,\omega,A,B,C_1,C_2,s\}$. Under such reparametrization, the observed Fisher information matrix~\eqref{eq:I_LPPLS_final} is transformed into
\begin{equation}\label{eq:I_zeta}
	I(\zeta)=J_D^TI_\eta(\eta(\zeta))J_D,
\end{equation}
where $J_D=\partial \eta/\partial\zeta$ is the Jacobian matrix of the transform from $\eta$ to $\zeta$,
whose its full expression is given by~\eqref{eq:jacobian_D} in Appendix C. Finally, the likelihood interval for 
the damping parameter is
\begin{equation}\label{eq:approx_LI_D}
	\mathrm{LI}(D;t_c)=\left\{D:|D-\hat D_{t_c}|<\Delta_{D;t_c}\right\},
	~\Delta_{D;t_c}=\sqrt{-2\ln c\left[I^{-1}(\hat\zeta_{t_c})\right]_{1,1}}.
\end{equation}
As discussed above, the modified profile likelihood~\eqref{eq:modified_lik} for the main parameter $t_c$ of interest is invariant 
with respect to such transformations of the nuisance parameter vector $\eta$.

We are now in position to discuss the overall results presented in Figure~\ref{fig:modified_profile}. 
The first important observation is that, in view of the determined likelihood intervals, the rejection of the ``suboptimal'' solution is no more 
warranted (given that $\mathrm{LI}(\omega)=\{5.47<\omega<6.07\}$). Observe that the optimal solution now easily fits in the extended interval of the damping parameter constraint~\eqref{eq:constrains}. Second, it is interesting to compare the interval widths ($2\Delta$) 
representing the uncertainty of the different parameters. In the particular example 
presented in Figure~\ref{fig:modified_profile}, the power law exponent $m$ is the most uncertain parameter with $2\Delta_{m;t_c}\approx 0.3$, which is about 40\% of the estimated value $\hat m_{t_c}$. The damping parameter $D$, which is proportional to $m$, also has a fairly broad likelihood interval with width $2\Delta_{D;t_c}\approx 0.17$, which is about 20\% of the estimated value $\hat D_{t_c}$. Finally, the uncertainty of the logperiodic angular frequency is $2\Delta_{\omega;t_c}\approx 0.63$, which is about 7\% of $\hat\omega_{t_c}$. 
In general, the widths $2\Delta$ of the likelihood intervals strictly depend on the specific realisation of the data, but our extensive tests
have shown that the above observations typically hold. Finally, it is interesting to document that such rather large uncertainty in the nuisance parameters does not result in a dramatic change of the likelihood intervals for the parameter of interest $t_c$. And while the modified profile likelihood corrects the shape of the distribution, the intervals~\eqref{eq:LI_tc} for the profile and modified profile likelihoods at a 5\% cutoff agree rather well in this and many another cases. 


\section{Application of the methodology}
\label{sec:examples}

In the previous Sections \ref{sec:likelihood}--\ref{sec:m_w}, we have developed a framework to infer the critical time $t_c$ from the LPPLS model, which includes
parameter estimation together with its confidence interval, as well as the confidence intervals of 
the relevant nuisance parameters \emph{within a fixed calibration window $[t_1,t_2]$}. However, for real life applications, one cannot limit oneself to the analysis of a single time-scale, because financial time-series result from complex generating processes,
from volatility clustering of the simplest form to multifractal models, subjected to regime-shifts leading to non-trivial scaling structures.
In order to understand the complexity of such phenomena through the prism of some model
like LPPLS, one needs to apply this model at different scales simultaneously, and also consider the evolution of the model parameters in time. 

In this section, we extend the analysis of the LPPLS model to the scale-domain $t_2-t_1$ and provide illustrations of the application of the methodology both to synthetic case and real price series.

\subsection{Aggregation of time-scales}\label{sec:multi-scale}

By time scale, we mean the width $\Delta_t \equiv t_2-t_2$ of the time window in which the analysis is performed.
The aggregation of analyses performed at different time-scales is not a trivial problem, whose difficulty starts with the mere computational complexity of non-linear models. Usually, the application of the model at several time scales proportionally increases the computational time, and the output data that needs to be analyzed also increases manifold. Further, in order to make the analysis operational, 
one needs a method for aggregating the massive amount of parameter information for the construction of the predictive features or signals. Then, the next step is to perform a full-scale back-testing of the constructed signals for understanding their predictive power. These challenging operational steps go beyond the scope of the present methodological paper, and will be reported elsewhere. Some practical aspects are already discussed in~\citep{Sornette_etal2015_Shanghai}, where multi-scale signals were used for ex-ante forecasting the crash in Chinese markets in June 2015. \cite{SornetteZhoufor2006} also presented a multi-scale analysis with LPPLS, in which the different scales
were combined via a pattern recognition algorithm. Here, we will focus on the descriptive analysis and visualization aspects.

The analysis of the modified profile likelihood in Sections~\ref{sec:modified}--\ref{sec:m_w} was aimed at estimating the likelihood intervals (LI) of the critical time $t_c$ as well as of the logperiodic angular frequency $\omega$, power law exponent $m$ and damping $D$, which 
contribute to the filtering criteria~\eqref{eq:constrains}. A multi-scale approach would require analysis of these outputs for different values of window sizes $\Delta_t=t_2-t_1$. One of the most natural ways is to construct the modified profile likelihood $L_m(t_c;\Delta_t)$ independently for different window sizes $\Delta_t$. Because the absolute value of the likelihood depends on the amount of data, it does not make sense to compare value of $L_m$ for different $\Delta_t$ directly.  For comparison, we will use the normalization as in~(\ref{eq:rel_lik}), and will apply it for each window size $\Delta_t$ independently, constructing the \emph{relative multi-scale modified profile likelihood} $R(t_c,\Delta_t)=L_m(t_c; \Delta_t) / \max_{t_c}L_m(t_c; \Delta_t)$. The structure of $R(t_c,\Delta_t)$ then directly provides with scale-dependent likelihood intervals $\mathrm{LI}(t_c;\Delta_t)$ for the critical time $t_c$. 

\begin{figure}[h!]
  \centering
  \includegraphics[width=\textwidth]{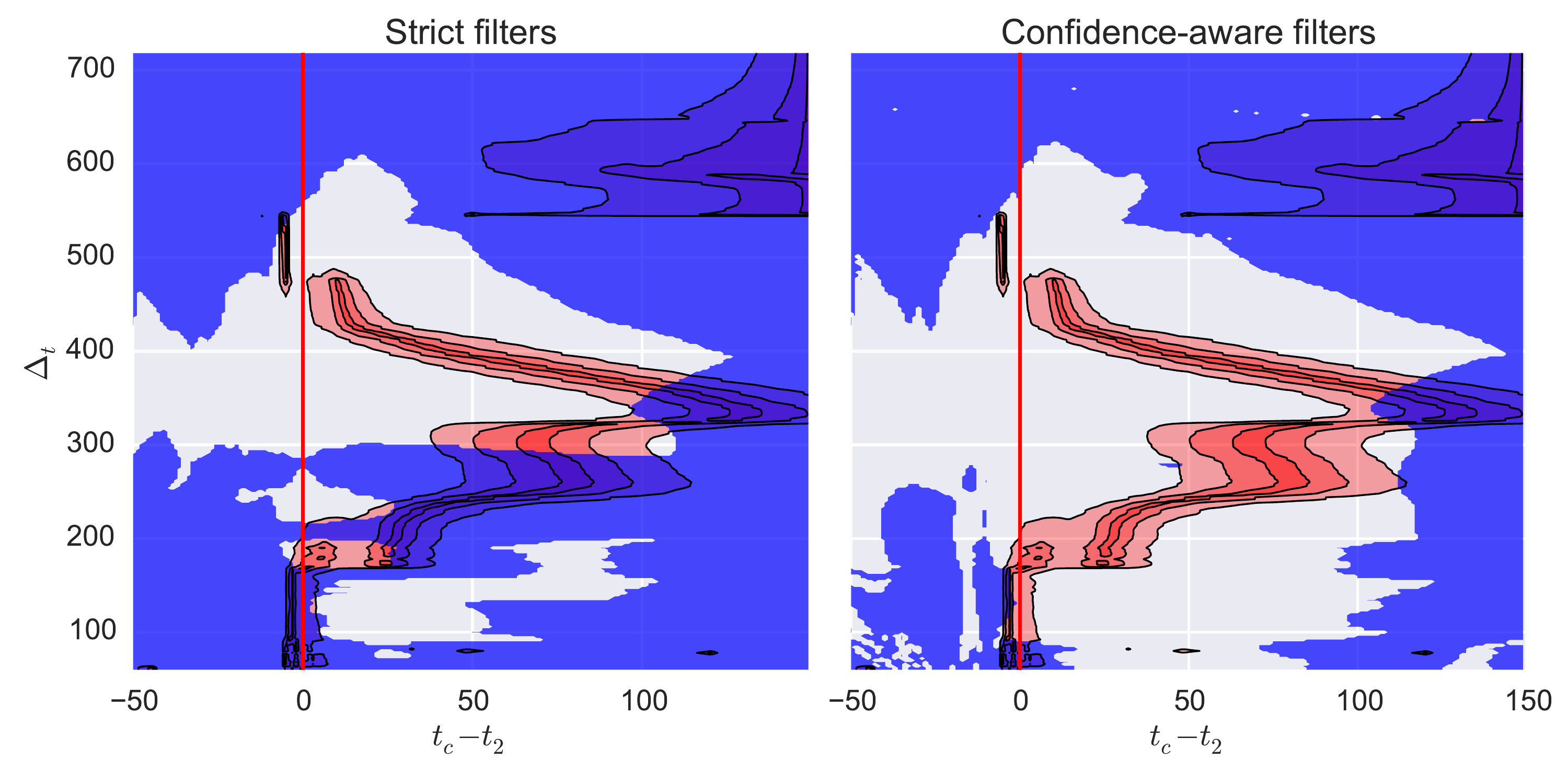}
  \caption{Two-dimensional structure of the relative multi-scale modified profile likelihood $R(t_c,\Delta_t)$ at the date $t_2$=2015-06-12, 
corresponding to multiple calibrations of the model with different window sizes $\Delta_t$. The horizontal axis gives the value $t_c-t_2$ 
with the solid red vertical line indicating the case where $t_c$ is coincident with the date of analysis. Each horizontal slice of the plot gives 
in color code the dependence of the individual modified profile likelihoods of the model (as in Figure~\ref{fig:profile_m_w}) calculated for a given window size $\Delta_t=t_2-t_1$ (vertical axis) as a function of $t_c-t_2$. The shaded red area corresponds to likelihood intervals $\mathrm{LI}(t_c)$ of the critical time at the 5\%, 50\% and 95\% cutoff (from lightest to darkest colors). The shaded blue area denotes values of $(t_c,\Delta_t)$, where the constraints on the nuisance parameters~\eqref{eq:constrains} are not met. The left panel corresponds to the case when only MLE parameters are considered for filtering, the right panel corresponds to the case when likelihood intervals~\eqref{eq:approx_LI_m_w} and~\eqref{eq:approx_LI_D} are taken into account.} 
  \label{fig:2d_profile}
\end{figure}

In order to illustrate this approach, we construct modified profile likelihoods for $\Delta_t$ varying from 60 to 700 days 
and for $t_c$ varying from $t_2-50$ to $t_2+150$ days. The scale-dependent likelihood intervals are presented in red color in Figure~\ref{fig:2d_profile} (red profiles are identical in both panels) for the same $t_2$=2015-06-12 that was used for illustration earlier in the paper. One can see the same bi-modal structure for $\Delta_t=180$ as reported earlier, which
suggests two possible scenarios for the end of the bubble: $t_c=t_2+5$ and $t_c=t_2+25$ (days). Figure~\ref{fig:2d_profile} gives an illustrative overview of the structure of the inferred critical time $t_c$ broken down in three time scales:
(i) short time scales ($\Delta_t\leq160$) suggest that the price trajectory is at its peak already and the critical time is close to the date of analysis $t_c\approx t_2$ (with the MLE of $t_c$ 
being a few days before $t_2$); (ii) intermediate time scales ($180\leq\Delta_t\leq350$) suggest two main scenarios 
in which the critical time is clustered around 20-30 or 60-90 days in the future; 
and (iii) large scales $\Delta_t>350$ do not give stable clusters.

Figure~\ref{fig:2d_profile} also provides important insights on the range of values $(t_c;\Delta_t)$ 
for which the parameters obey the theoretical constraints~\eqref{eq:constrains} --- below we will refer to them as \emph{qualified fits}. In the left panel, 
the blue shaded area indicates when LPPLS fits can be rejected based solely on the MLE values of $m,\omega$ and $D$ (``strict filtering''). 
In the right panel, we take additionally into account the likelihood intervals of these parameters (see Section~\ref{sec:m_w}) and 
show only the region where these intervals have no overlap with the constraints~\eqref{eq:constrains} (``confidence-aware filtering''). 
These cases differ quite dramatically, in the sense that strict filtering falsely rejects a substantial number of fits that correspond to
credible alternative scenarios. As discussed in~\citep{Sornette_etal2015_Shanghai}, choosing the proper filters is one of
the key ingredients for constructing successful signals. Being a very broad subject, 
constructing and testing useful filters goes beyond the scope of present paper.
For the time being, we stress how crucial it is to take into account data-induced uncertainty when 
constructing signal filters.

Let us now describe some of the potential numerical issues that often arise in such complicated optimization problems. 
First, because the search for $t_c$ is constrained in a pre-defined bounded interval, the real $\max_{t_c}L_m(t_c; \Delta_t)$ might 
lie outside of it, so that the numerical procedure might pick up a value at the boundary of the search space on $t_c$. Normalising
$L_m(t_c)$ to $1$ at this boundary point, this may result in having a wide range of high values of $R(t_c,\Delta_t)$ close to this boundary,
leading to a spurious likelihood interval $\mathrm{LI}(t_c)$. In the example above, this is exactly
what happens for $\Delta_t>550$ (red profiles at the top-right of Figure~\ref{fig:2d_profile}), where the maximum of the modified profile likelihood is beyond the search range ($t_c>t_2+150$) and the inference on the likelihood intervals is completely misleading.

Another problem is the potential bad convergence of the optimization of the nuisance parameters in~\eqref{eq:profile}, which dramatically affects the value of $L_p(t_c)$ and thus $L_m(t_c)$. Usually, this situation occurs for large values of $t_c-t_2$, especially
when $t_c-t_2$ is not small compared with the window size $\Delta_t$. However, it highly depend on the  structure of the residuals and the numerical method might not converge even for moderate values of $t_c-t_2$. What makes this issue complicated is that there is no simple way of detecting bad convergence, neither algorithmically nor even visually in plots like Figure~\ref{fig:2d_profile}. It often results in some kind of discontinuities in the plot, but not always. Take for instance the case $\Delta_t=470-480$, where an apparent discontinuity of likelihood intervals is in fact the consequence of a continuous transition of one maximum of the likelihood to another when increasing $\Delta_t$. 

As with all non-linear optimization problems, there is no ``silver bullet'' to address such numerical issues.  Measures such as increasing the region of search or the precision of numerical methods do not always help. Especially when one performs fully automated analyses, it is highly recommended to carefully validate each step of the procedure and take outputs with a grain of salt, not hesitating to ``triple-check'' any suspicious results.

\subsection{Synthetic tests}\label{sec:synthetic}

\begin{figure}[h!]
  \centering
  \includegraphics[width=0.95\textwidth]{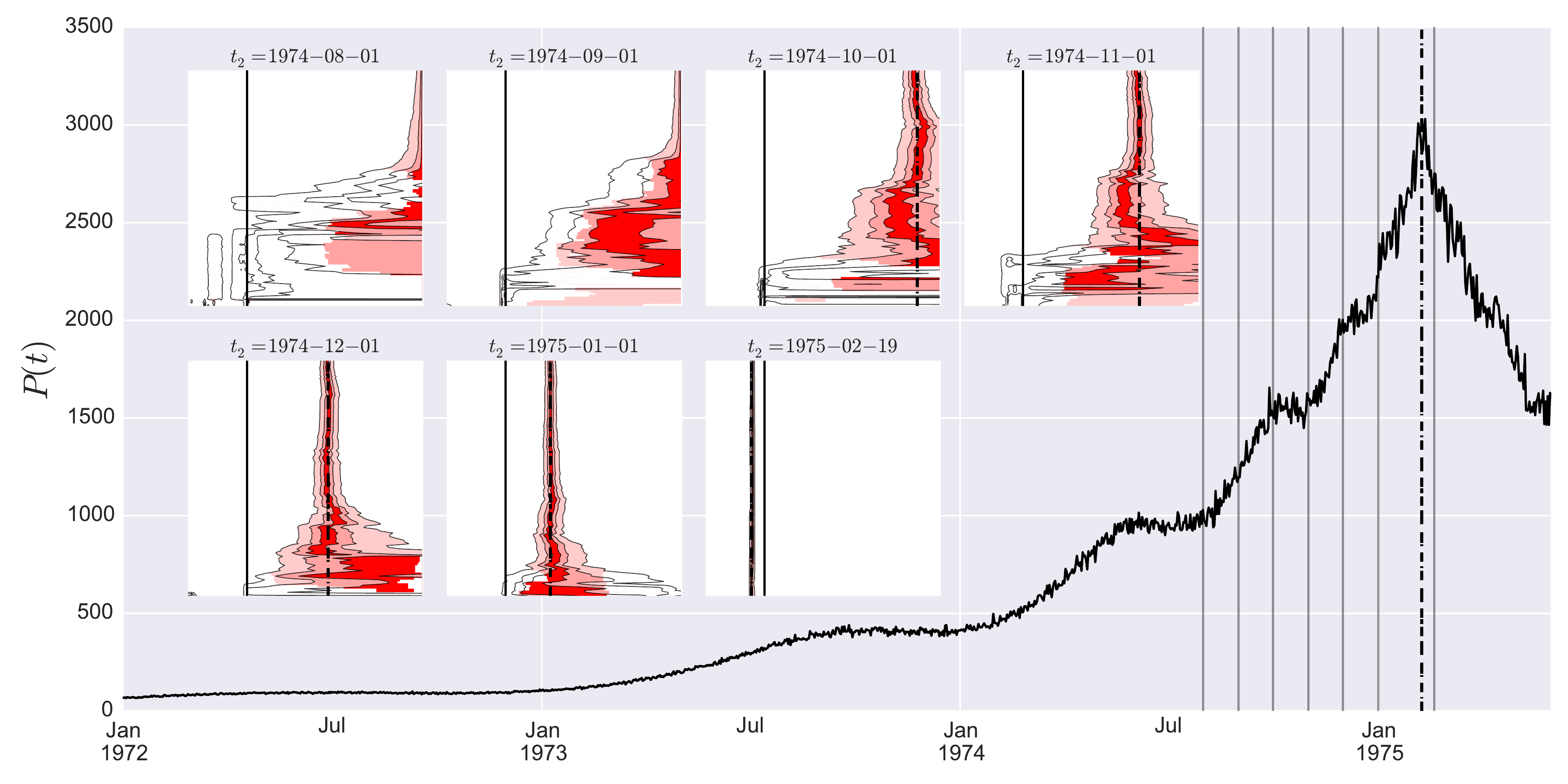}
  \caption{Synthetic price time series~\eqref{eq:syn_lppl} together with multi-scale modified profile likelihoods calculated at various dates $t_2$. Each inset shows contour plots of the likelihood intervals at 5\%, 50\% and 95\% cutoff levels (as in Figure~\ref{fig:2d_profile}). The red shaded area denotes values of $(t_c, \Delta_t)$ where the constraints on the nuisance parameters~\eqref{eq:constrains} are met when the likelihood intervals~\eqref{eq:approx_LI_m_w} and~\eqref{eq:approx_LI_D} are taken into account (i.e.\ red color denote these parts of the contour plots that are not covered by blue area in Figure~\ref{fig:2d_profile}). The solid vertical line corresponds to $t_c=t_2$, and the dashed vertical line shows the true critical time $t_c=t_c^{0}$. Values of $t_2$ used for the analyses are indicated in the inset titles. They are also shown with vertical gray lines in the plot with the price trajectory. } 
  \label{fig:contour_lppls}
\end{figure}

In order to gain insight about the likelihood inference of the critical time ($t_c$) during a growing bubble and establish a solid background for our empirical analysis, we first test our methodology on synthetic time series, where the underlying process follows the LPPLS structure. Specifically, we generate the log-price as
\begin{equation}\label{eq:syn_lppl}
	\ln[P(t)] = LPPLS(t) +\sigma\epsilon(t),
\end{equation}
where $LPPLS(t)$ is given by~\eqref{eq:lppl_symmetry} with $t_c^{0}$=1975-02-09, $m^{0}=0.8$, $\omega^{0}=9$, $\phi^{0}=0$ and $A^{0}=8$, $B^{0}=-0.015$, $C^{0}=0.0015$ (i.e. with low damping $D^{0}=0.88$), $\epsilon(t)$ is an iid N(0,1) noise and $\sigma^{0}=0.03$. The resulting price trajectory is illustrated in Figure~\ref{fig:contour_lppls}.

With the goal of understandings the evolution of the parameters as a function of the ``present'' time $t_2$ (the time of analysis)
for such a synthetic bubble, we apply our methodology to construct a multi-scale modified profile likelihood (see Section~\ref{sec:multi-scale} and Figure~\ref{fig:2d_profile}) at different dates $t_2$ increasing towards the end of the bubble at $t_c^{0}$. The resulting multi-scale profiles are shown in insets of Figure~\ref{fig:contour_lppls}. Far from the critical time (insets 1 and 2: $t_c^{0}-t_2=192$ and 161 days respectively), the critical time cannot be identified even in such a clean synthetic case with weak noise (partially because we are limiting the search space to $t_c<t_2+150$ days). Even when $t_c^{0}-t_2$ enters the range $<150$, the parameters continue to 
exhibit a large uncertainty. One can observe that fits for different scales progressively build a consensus,
as the likelihood peaks aggregate around the true critical time with a narrow likelihood interval around it. 
This is first observed for the large scales $\Delta_t>300-400$ (insets 3 and 4 for $t_c^{0}-t_2=131$ and 100 days respectively), and 
this consensus spreads to smaller scales of $\Delta_t\sim100-300$ (insets 5 and 6 for $t_c^{0}-t_2=70$ and 39 days respectively).

Another remarkable fact is that, once the critical time is passed and the price trajectory switched to a crashing regime ($t_2>t_c^0$, inset 7), all scales confirm this occurrence by fixing the MLE $\hat t_c\approx t_c^0$ with an extremely narrow likelihood interval, and this 
anchoring holds for a large time interval. The same effect is observed in the analysis of real data 
presented in Section~\ref{sec:casestudies} --- even when the ex-ante forecast of the end of the bubble might be difficult or inconclusive, the change of the price direction can be identified quite reliably within a few days of the switching point.

\subsection{Case-studies}\label{sec:casestudies}

We now provide examples of the application of the procedure described
in previous sections to several well-known historical bubbles: (i) the rally in the US markets in the second half of the 1980's culminating with the Black Monday crash of Oct. 19. 1987, (ii) the dot-com bubble in the IT sector in the US culminating with a crash in April 2000, (iii) the Chinese bubble of 2014-2015 that peaked in June 2015.

We use the daily closing prices of the S$\&$P 500, NASDAQ and SSEC indices provided by Thomson Reuters Dataworks Enterprise (DWE). We only consider business days, ignoring weekends and one-day holidays. However, for extended holidays (such as the Chinese New Year in 2015, when exchanges were closed over February 7-13), we fill the gaps with the closing price of the previous day. For calibrations using business time (see Section~\ref{sec:lppls}), such data preprocessing would not be necessary.

\begin{figure}[!tp]
 \centering
\includegraphics[width=0.95\textwidth]{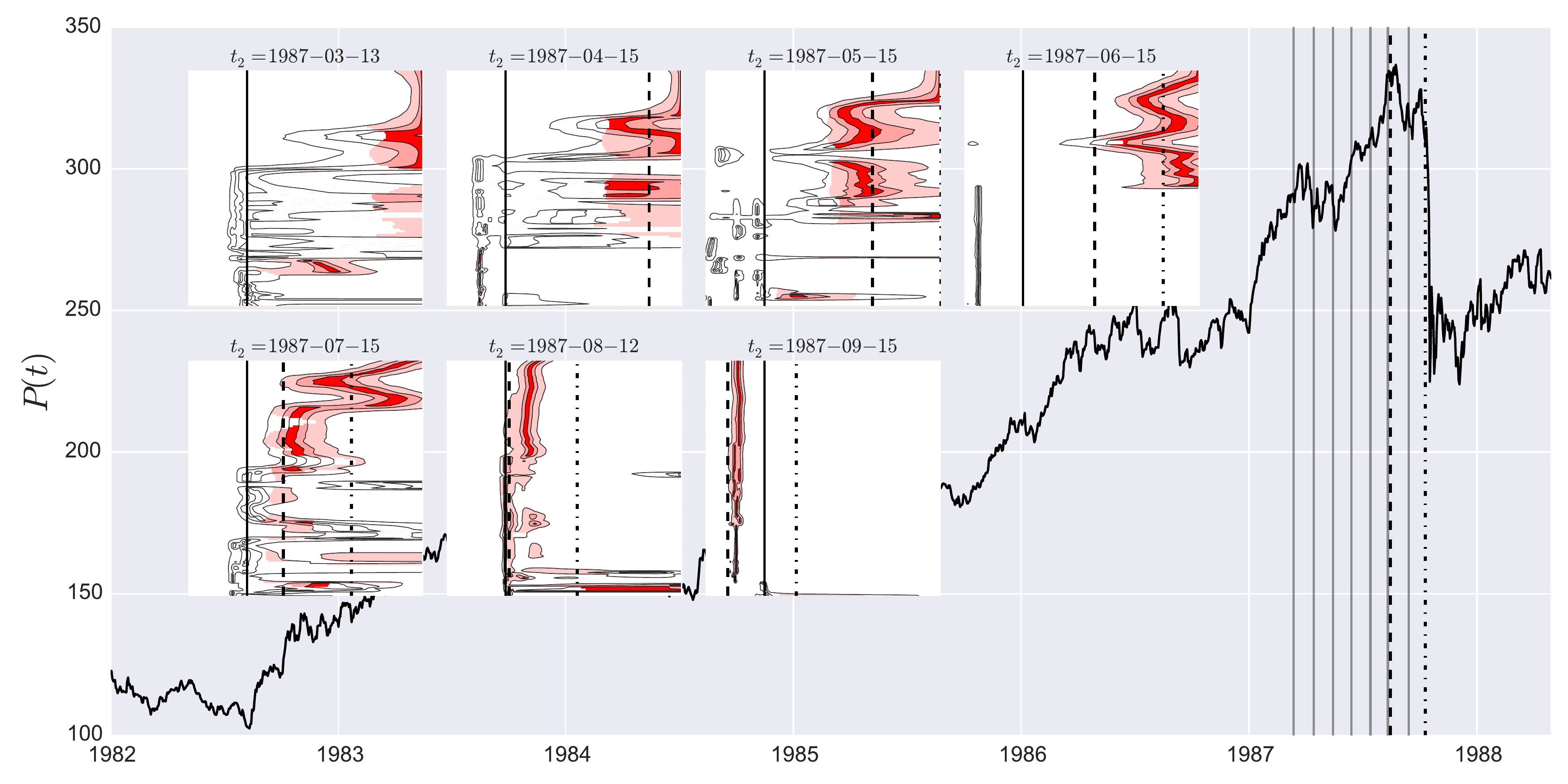}
\caption{Same as Figure \ref{fig:contour_lppls}  for the $S\&P 500$ index that shows a strong bubble of US markets developing in the second half of the 1980's, which culminated with the Black Monday crash of Oct. 19. 1987.} 
\label{fig_sp500gil}
\end{figure}

\begin{figure}[!tp]
 \centering
\includegraphics[width=0.95\textwidth]{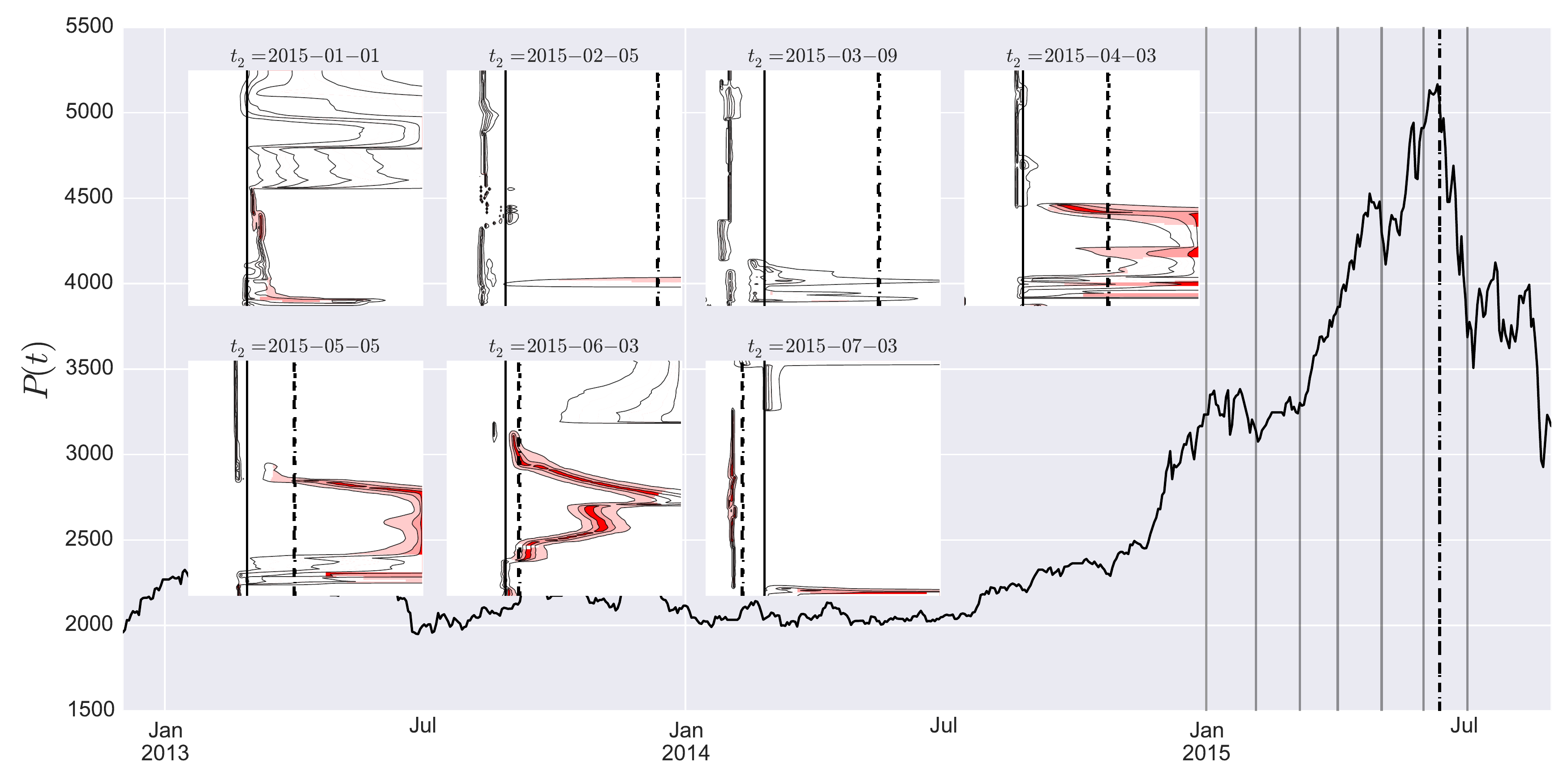}
\caption{Same as Figure \ref{fig:contour_lppls}  for the Chinese bubble of 2014-2015 that peaked in June 2015.} 
\label{fig_ssecgil2}
\end{figure}

\begin{figure}[!tp]
 \centering
\includegraphics[width=0.95\textwidth]{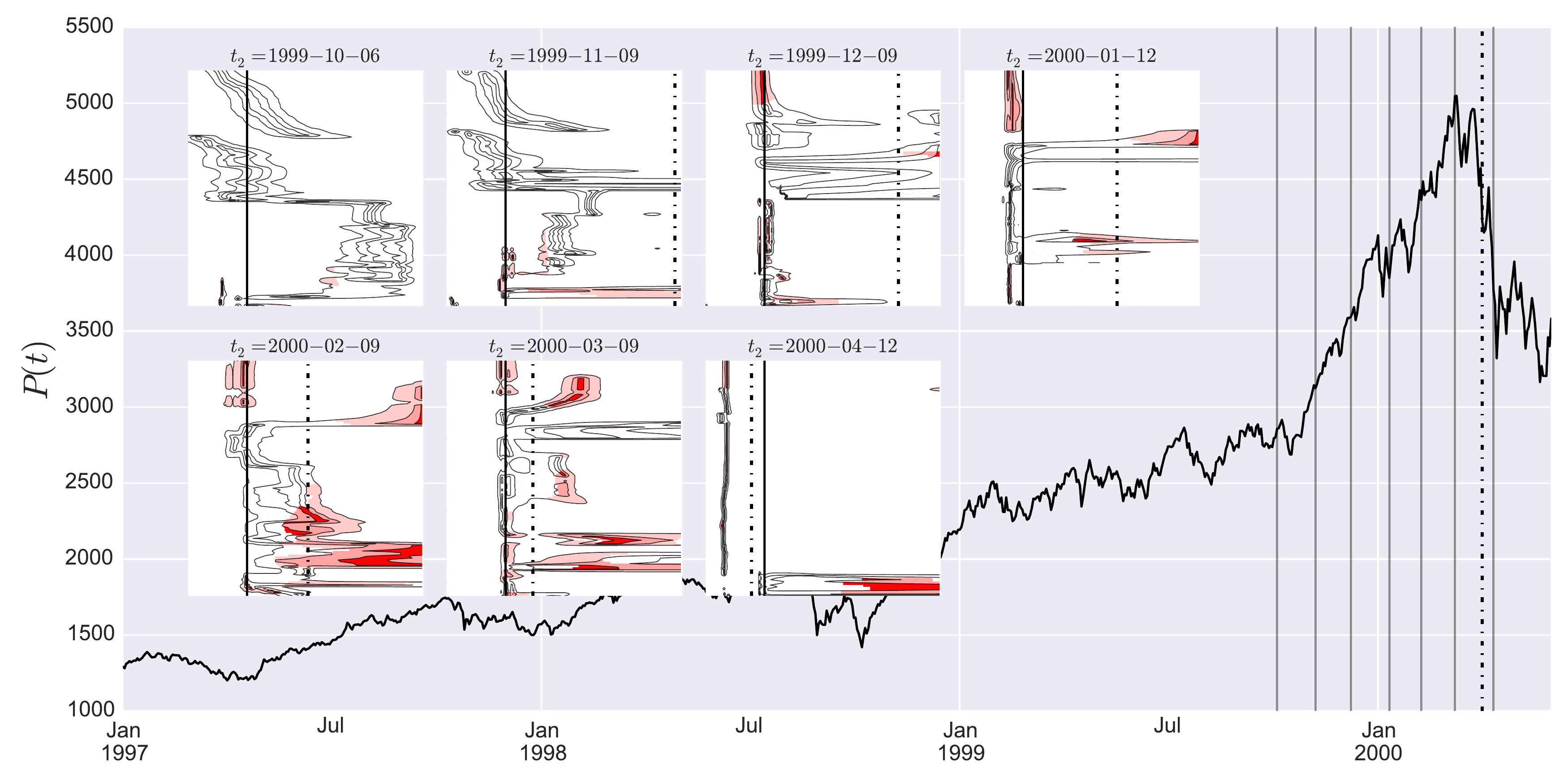} 
\caption{Same as Figure \ref{fig:contour_lppls}  for the dot-com bubble in the IT sector in the US culminating with a crash in April 2000 } 
\label{fig_nasdaqgil}
\end{figure}

Employing the procedure explained in Sec. \ref{sec:synthetic}, we obtain Figures \ref{fig_sp500gil}-\ref{fig_nasdaqgil}.
In each of these three figures, the main graph shows the price time series $P(t)$ together with vertical dashes lines that identify remarkable
turning points of the price dynamics. In the case of the S$\&$P 500, we show two different vertical dashed lines
associated with the two peaks of the index preceding the crash. The seven thin vertical lines indicate
the position of the seven $t_2$ values chosen for the construction of the Likelihood intervals of $t_c$.
The seven insets show contour plots of the likelihood intervals at 5\%, 50\% and 95\% cutoff levels (as in Figure~\ref{fig:contour_lppls}).
 
Figures \ref{fig_sp500gil} for the S\&P 500 shows that the Profile Likelihhod of $t_c$ as a function of time scale $\Delta_t \equiv t_2-t_1$
and ``present time'' $t_2$ is very similar to those obtained in synthetic tests. 
As early as $t_2$ = 1987-04-15, one can visualize the high Likelihood of $\widehat{t}_c \approx Oct.~1987$ over almost all time scales.
Interestingly, the Likelihood interval narrows down as $t_2$ approaches the end of the bubble. Moreover, there is an increase
of the number of qualified fits (those where constraints on the model parameters~\eqref{eq:constrains} are met when the likelihood intervals~\eqref{eq:approx_LI_m_w} and~\eqref{eq:approx_LI_D} are taken into account --- shown as the red-shaded region) at $t_2$ increases. These two results can be rationalized
by the fact that more information relevant to the identification of the bubble become available as more data are used.

As shown in figure \ref{fig_ssecgil2}, similar observations carry over to the SSEC bubble ending in June 2015, albeit
with a smaller number of qualified fits. One can observe that the analyses performed for the time scales
$\Delta_t \in [400,500]$ and $\Delta_t \in [100,200]$ provide a correct diagnostic of the end of the bubble with 
a narrow confidence interval. The time scale $\Delta_t \in [400,500]$ correctly locks in on the true peak as early as April 2015. 

Figure \ref{fig_nasdaqgil} shows the same analysis for the dotcom bubble that developed in the NASDAQ Index.
At $t_2$ = 2000-02-09, the time scales $\Delta_t \in  [100,350]$ correctly lock in on the true peak $\approx$ April 2000.
The other intermediate time scales give an estimation $\widehat{t}_c$ of the end of the bubble that agrees
with the empirical value within the 95\% confidence of the likelihood intervals.. 
All estimates on different $t_2$'s appear to be either unqualified or signalling a different value for the change of regime to occur.

Overall, these empirical results exhibit the following behaviors:
(i) for $t_2$ far from $t_c$, there are fewer qualified fits and different scales tend to provide
distinct estimates $\widehat{t}_c$; (ii) when approaching the true $t_c$, the Likelihood intervals for $\widehat{t}_c$
start to align, with the formation of clusters associated with different possible scenarios; (iii) rather close to the
true $t_c$, one can often observe a strong cluster around $\widehat{t}_c \approx t_2$ and a narrow likelihood interval.

On the other hand, the fact that different time scales used for fitting the LPPLS model tend to 
suggest different values of $\hat{t}_c$ is important to keep in mind, as 
this observation is in contrast with the behavior obtained for synthetic time-series. 
This is likely due to ``model error'', i.e., the simple LPPLS model (\ref{eq:lppl})
is only an approximation of the unknown true generating process of the price dynamics. For instance, 
earlier works \citep{JohansenSornette1997Lan,JohansenSornette1999,Sornette2002fractal_functions,ZhouSorRen03} 
have pointed out the important of including higher harmonics and more complex forms
generalising this simple first-order LPPLS formula (\ref{eq:lppl}).

We thus stress the importance of employing filtering criteria to decrease the probability of the occurrence
of errors of type I (``false positives''). The Likelihood Method has been shown to provide more reliable interval estimates for the critical time
than simple OLS point estimates, in particular as $t_2$ approaches $t_c$.

\section{Concluding remarks \label{sec:conclusion}}

We have presented a detailed methodological study of the application of the modified profile likelihood method 
proposed by~\citet{BarndorffNielsen1983}, with the goal of tackling the instabilities and uncertainties occurring in the calibration
of nonlinear financial models characterised by a large number of parameters. We have taken the Log-Periodic Power Law Singularity 
(LPPLS) model as an example for the application of the methodology.
This is motivated by the claims of the LPPLS model to provide useful estimations of the end of bubbles and their crashes,
which can be interpreted as critical times $t_c$.  One of our major advances has been to formulate the calibration procedure 
in a way such that the critical time $t_c$ of a given bubble becomes the major parameter of interest in the likelihood inference.
In contrast, the other model parameters are treated as nuisance parameters.  
While the problem of dealing with nuisance parameters is 
not new in Statistics, the present article is, to our knowledge, the first one in quantitative finance that elaborate in details how 
to deal with them to obtain better inference on the target parameter (here $t_c$). We have shown that it is possible to bypass the strong nonlinearity of the model
by using a very precise approximation for the modified profile likelihood. 
This has allowed us to provide a systematic construction of the 
parameter estimation uncertainties and of the corresponding likelihood intervals, both 
for the target parameter $t_c$ and for the other so-called nuisance parameters.
We have also introduced the importance of performing the calibrations at multiple time scales,
i.e., in time windows of many different sizes typically from 100 to 750 days.
This has led us to provide representations to aggregate the results obtained from 
the calibrations at different time scales, thus obtaining a multi-scale picture of the
possible scenarios for the development of on-going bubbles. We have tested the methodology
on synthetic price time series and on three well-known historical financial bubbles.

\section*{Acknowledgments}
We are grateful to Professor Thomas A.\ Severini for helpful discussions about the approximation of the modified profile likelihood function. We also thank Diego Ardila Alvarez for many fruitful discussions while preparing the manuscript.

The analysis in the paper was performed using open source software: Python 2.7 (\url{http://www.python.org}) and libraries: Pandas~\citep{McKinney_Pandas2012}, NumPy~(\url{http://www.numpy.org/}), SciPy~(\url{http://www.scipy.org/}), IPython~\citep{Perez_IPython2007}, Matplotlib~\citep{Hunter_Matplotlib2007} and Seaborn~(\url{https://www.stanford.edu/~mwaskom/software/seaborn}).

\appendix
\section{Derivation of the approximated modified profile likelihood for the LPPLS model}\label{app:derive}

We derive the approximated expression for the modified profile likelihood~\eqref{eq:modified_lik_approx} of the LPPLS model~\eqref{eq:lppl2}. The parameter of interest is the critical time $t_c$ and nuisance parameters $\eta=\{\psi,s\}$ include both the vector $\psi=\{m,\omega,A,B,C_1,C_2\}$ of other LPPLS parameters and the variance $s$ of the error term. 

\vspace{-0.4cm}
\subsection{Observed Fisher information}

The calculation of the observed Fisher information matrix $I(\hat\eta_{t_c})$ is straightforward. According to~\eqref{eq:observed_I}, 
it can be written in the form of a block matrix:
\begin{equation}\label{eq:I_LPPLS}
	I(\hat\eta_{t_c})=-\begin{pmatrix}
	\ell_{\psi,\psi}(\hat\eta_{t_c}) & \ell_{\psi,s}(\hat\eta_{t_c})\\
	\ell^T_{s,\psi}(\hat\eta_{t_c}) & \ell_{s,s}(\hat\eta_{t_c})\\
	\end{pmatrix},
\end{equation}
where $\ell_{\cdot,\cdot}(\hat\eta_{t_c})$ denotes the respective second partial derivatives of the log-likelihood $\ln L(t_c,\psi,s)$ evaluated at the point $\hat\eta_{t_c}=\{\hat\psi_{t_c},\hat s_{t_c}\}$. For the likelihood~\eqref{eq:lik}, the first partial derivatives are given by:
\begin{equation}\label{eq:partial_1}
	\begin{array}{rl}
		\displaystyle  \frac{\partial \ell}{\partial \psi}&=
		\displaystyle-\frac1{2s}\frac{\partial \sse(t_c,\psi)}{\partial \psi}; \\
		\displaystyle  \frac{\partial \ell}{\partial s}&=
		\displaystyle-\frac{n}{2s}+\frac{\sse(t_c,\psi)}{2s^2}~.\\
	\end{array}
\end{equation}
In turn, the second partial derivatives read:
\begin{equation}\label{eq:partial_2}
	\begin{array}{rl}
		\displaystyle  \frac{\partial^2 \ell}{\partial \psi\partial \psi^T}&=
		\displaystyle -\frac1{2s}\frac{\partial^2 \sse(t_c,\psi)}{\partial \psi\partial \psi^T}; \\
		\displaystyle  \frac{\partial^2 \ell}{\partial \psi\partial s}&=
		\displaystyle \frac1{2s^2}\frac{\partial \sse(t_c,\psi)}{\partial \psi}; \\
		\displaystyle  \frac{\partial^2 \ell}{\partial s^2}&=
		\displaystyle \frac{n}{2s^2}-\frac{\sse(t_c,\psi)}{s^3}.\\
	\end{array}
\end{equation}
The MLE $\hat\eta_{t_c}$ is given by the global maximum of $\ln L(t_c,\eta)$ for fixed $t_c$, so $\hat\psi_{t_c}$ is given by a global minimum of $\sse(t_c,\psi)$, thus:
\begin{equation}\label{eq:l_psi_s}
	\ell_{\psi,s}(\hat\eta_{t_c})=
	\left. \frac{\partial^2 \ell}{\partial \psi\partial s}\right|_{\eta=\hat\eta_{t_c}}=
	\frac1{2\hat s_{t_c}^2}\left.\frac{\partial \sse(t_c,\psi)}{\partial \psi}\right|_{\psi=\hat\psi_{t_c}}
	=\Theta,
\end{equation}
where $\Theta=\{0,0,0,0,0,0\}^T$ is the vector of zeros.
Taking into account~\eqref{eq:est_s_tc}, we can write for the third term in~\eqref{eq:partial_2}:
\begin{equation}\label{eq:l_s_s}
	\ell_{s,s}(\hat\eta_{t_c})=
	\left. \frac{\partial^2 \ell}{\partial s^2}\right|_{\eta=\hat\eta_{t_c}}=
	\frac{n}{2\hat s_{t_c}^2}-\frac{\sse(t_c,\hat\psi_{t_c})}{\hat s_{t_c}^3}=-\frac{n}{2\hat s_{t_c}^2}.
\end{equation}
Finally, plugging~\eqref{eq:l_psi_s} and~\eqref{eq:l_s_s} into~\eqref{eq:I_LPPLS}, we obtain the following form for the observed Fisher information:
\begin{equation}\label{eq:I_LPPLS_final}
	I(\hat\eta_{t_c})=\begin{pmatrix}
	\displaystyle \frac1{2\hat s_{t_c}}
	\left.\frac{\partial^2 \sse(t_c,\psi)}{\partial \psi\partial \psi^T}\right|_{\psi=\hat\psi_{t_c}} 
	& \Theta \\ \Theta^T &
	\displaystyle \frac{n}{2\hat s_{t_c}^2}
	\end{pmatrix}
	=\begin{pmatrix}
	\displaystyle \frac1{\hat s_{t_c}}
	\sum_{i=1}^n\left.\frac{\partial^2 \lppls(\tau_i;t_c,\psi)}{\partial \psi\partial \psi^T}\right|_{\psi=\hat\psi_{t_c}} 
	& \Theta \\ \Theta^T &
	\displaystyle \frac{n}{2\hat s_{t_c}^2}
	\end{pmatrix},
\end{equation}
and its determinant
\begin{equation}\label{eq:det_I_LPPLS}
	|I(\hat\eta_{t_c})|= 
	\frac n{2\hat s_{t_c}^{p+2}}
	\left|\sum_{i=1}^n\frac{\partial^2 \lppls(\tau_i;t_c,\psi)}{\partial \psi\partial \psi^T}\right|_{\psi=\hat\psi_{t_c}},	
\end{equation}
where $p=\dim\psi=6$.

\subsection{Covariance matrix}

Here, we calculate the covariance matrix $\Sigma(t_c,\hat\eta_{t_c};\hat t_c, \hat\eta)$~\eqref{eq:covariance}. For this, we will first evaluate the general form of the matrix~\eqref{eq:covariance} and then substitute $(t_c,\hat\eta_{t_c})$ and $(\hat t_c, \hat\eta)$. Similarly to the Fisher information, the matrix $\Sigma$~\eqref{eq:covariance} can be written in a block form:
\begin{equation}\label{eq:sigma_LPPLS}
	\Sigma\left(t_{c;1},\eta_{1};t_{c;2},\eta_{2}\right)=
	\mathrm{E}_{(2)}\left[\begin{pmatrix}
	\ell_{\psi}(1)\ell^T_{\psi}(2) &
	\ell_{\psi}(1)\ell_{s}(2) \\
	\ell_{s}(1)\ell^T_{\psi}(2) &
	\ell_{s}(1)\ell_{s}(2) \\
	\end{pmatrix}\right],
\end{equation}
where $\ell_\cdot(1)$ symbolically denotes the first partial derivative~\eqref{eq:partial_1} of the log-likelihood evaluated at~$\left(t_{c;1},\eta_{1}\right)=\left(t_{c;1},\psi_{1},s_1\right)$; similarly, $\ell_\cdot(2)$ is evaluated at~$\left(t_{c;2},\eta_{2}\right)=\left(t_{c;2},\psi_{2},s_2\right)$. Given~\eqref{eq:SSE}, the partial derivative of the SSE that enters~\eqref{eq:partial_1} has the form:
\begin{equation}\label{eq:partial_SSE}
	\frac{\partial\sse(t_c,\psi)}{\partial\psi}
	=-2\sum_{i=1}^n\varepsilon_i\frac{\partial \lppls_i}{\partial\psi},
\end{equation}
where we have denoted $\varepsilon_i=\varepsilon(\tau_i;t_c,\psi)$ and $\lppls_i= \lppls(\tau_i;t_c,\psi)$.

Let us first consider the cross-terms in~\eqref{eq:sigma_LPPLS}. We substitute~\eqref{eq:partial_SSE} into~\eqref{eq:partial_1} and then into~\eqref{eq:sigma_LPPLS}. Then, after replacing the product of sums with the double sum and using the linearity of the expectation operation, we have:
\begin{equation}\label{eq:e_psi_s}
	\mathrm{E}_{(2)}[\ell_{\psi}(1)\ell_{s}(2)]=
	-\frac{n}{2s_1s_2}\sum_{i=1}^n \mathrm{E}_{(2)}[\varepsilon_i]
	\left.\frac{\partial \lppls_i}{\partial\psi}\right|_{\substack{t_c=t_{c;1}\\\psi=\psi_{1}}}
	+\frac1{2s_1s_2^2} \sum_{i=1}^n \sum_{j=1}^n
	\mathrm{E}_{(2)}[\varepsilon_i\varepsilon_j^2]
	\left.\frac{\partial \lppls_i}{\partial\psi}\right|_{\substack{t_c=t_{c;1}\\\psi=\psi_{1}}}.
\end{equation}
As discussed in Section~\ref{ssec:approx_modified}, the expectations in~\eqref{eq:e_psi_s} are taken with respect to the probability distribution that corresponds to the parameters $\{t_{c;2},\eta_{2}\}$, in other words under the assumption that $\varepsilon\sim N(0,s_{2})$. Thus $\mathrm{E}_{(2)}[\varepsilon_i]=\mathrm{E}_{(2)}[\varepsilon_i\varepsilon_j^2]=0$, and the cross-term is equal to the zero-vector:
$\mathrm{E}_{(2)}[\ell_{\psi}(1)\ell_{s}(2)]=\Theta$. Similarly for the second cross-term of~\eqref{eq:sigma_LPPLS}:
\begin{equation}\label{eq:e_psi_s_final}
	\mathrm{E}_{(2)}[\ell_{s}(1)\ell^T_{\psi}(2)]
	=\big(\mathrm{E}_{(2)}[\ell_{\psi}(1)\ell_{s}(2)]\big)^T=\Theta^T.
\end{equation}

Let us now consider the second derivatives with respect to the variance parameter $s$. Proceeding in the same way as above, we obtain:
\begin{equation}\label{eq:e_s_s}
	\mathrm{E}_{(2)}[\ell_{s}(1)\ell_{s}(2)] = \frac{n^2}{4s_1s_2}
	-\frac{n}{4s_1^2s_2}\sum_{i=1}^n \mathrm{E}_{(2)}[\varepsilon_i^2]
	-\frac{n}{4s_1s_2^2}\sum_{j=1}^n \mathrm{E}_{(2)}[\varepsilon_j^2]
	+\frac1{4s_1^2s_2^2}\sum_{i=1}^n \sum_{j=1}^n \mathrm{E}_{(2)}[\varepsilon_i^2\varepsilon_j^2].
\end{equation}
Taking into account that $\mathrm{E}_{(2)}[\varepsilon_i^2]=s_2$, $\mathrm{E}_{(2)}[\varepsilon_i^4]=3s_2^2$ and $\mathrm{E}_{(2)}[\varepsilon_i^2\varepsilon_j^2]=s_2^2$ (when $i\ne j$), we obtain:
\begin{equation}\label{eq:e_s_s_final}
	\mathrm{E}_{(2)}[\ell_{s}(1)\ell_s(2)]=\frac{n}{2s_1^2}.
\end{equation}
Finally, the submatrix term reads:
\begin{align}\label{eq:e_psi_psi}
	\mathrm{E}_{(2)}[\ell_{\psi}(1)\ell_{\psi}^T(2)] &= \frac1{s_1s_2}
	\sum_{i=1}^n \sum_{j=1}^n \mathrm{E}_{(2)}[\varepsilon_i\varepsilon_j]
	\left.\frac{\partial \lppls_i}{\partial\psi}\right|_{\substack{t_c=t_{c;1}\\\psi=\psi_{1}}}
	\left.\frac{\partial \lppls_j}{\partial\psi^T}\right|_{\substack{t_c=t_{c;2}\\\psi=\psi_{2}}}
	\nonumber\\
	&=\frac1{s_1}\sum_{i=1}^n 
	\left.\frac{\partial \lppls_i}{\partial\psi}\right|_{\substack{t_c=t_{c;1}\\\psi=\psi_{1}}}
	\left.\frac{\partial \lppls_i}{\partial\psi^T}\right|_{\substack{t_c=t_{c;2}\\\psi=\psi_{2}}},
\end{align}
where we have accounted for the fact that $\mathrm{E}_{(2)}[\varepsilon_i\varepsilon_j]=0$ when $i\ne j$.

The final expression is obtained by combining~\eqref{eq:e_psi_s_final},\eqref{eq:e_s_s_final} and~\eqref{eq:e_psi_psi} to~\eqref{eq:sigma_LPPLS} and evaluating the result at $(t_c,\hat\eta_{t_c};\hat t_c, \hat\eta)$:
\begin{align}\label{eq:sigma_LPPLS_final}
	|\Sigma(t_c,\hat\eta_{t_c};\hat t_c, \hat\eta)|&=
	\begin{vmatrix}
	\displaystyle \frac1{\hat s_{t_c}}\sum_{i=1}^n 
		\left.\frac{\partial \lppls_i}{\partial\psi}\right|_{\substack{t_c=t_{c}\\\psi=\hat\psi_{t_c}}}
		\left.\frac{\partial \lppls_i}{\partial\psi^T}\right|_{\substack{t_c=\hat t_{c}\\\psi=\hat\psi}}
	& \Theta \\ \Theta^T &
	\displaystyle\frac{n}{2\hat s_{t_c}^2}
	\end{vmatrix}\nonumber\\
	&=\frac{n}{2\hat s_{t_c}^{2+p}}\left| \sum_{i=1}^n 
		\left.\frac{\partial \lppls_i}{\partial\psi}\right|_{\substack{t_c=t_{c}\\\psi=\hat\psi_{t_c}}}
		\left.\frac{\partial \lppls_i}{\partial\psi^T}\right|_{\substack{t_c=\hat t_{c}\\\psi=\hat\psi}}
	\right|,
\end{align}
where $p=\dim\psi=6$. Note that a similar expression presented in~\cite{Severini1999} contains a typographical error in the power of the variance term.

\section{Partial derivatives of the LPPLS function}\label{app:lppls_derivs}

We present here the analytical expressions of the first and second partial derivatives of the LPPLS function~\eqref{eq:lppl2},
which are necessary for the calculation of the modified profile likelihood~\eqref{eq:Lm_LPPLS}--\eqref{eq:H}.

The first-order derivatives have the following forms:
\begin{equation}\label{eq:LPPLS_partial_1}
	\begin{array}{rl}
		\partial\lppls / \partial m&=|t_c-t|^m\ln|t_c-t|\Big[
		B+C_1\cos\big(\omega\ln|t_c-t|\big)+C_2\sin\big(\omega\ln|t_c-t|\big)\Big];\\
		\partial\lppls / \partial \omega&=|t_c-t|^m\ln|t_c-t|\Big[
		-C_1\sin\big(\omega\ln|t_c-t|\big)+C_2\cos\big(\omega\ln|t_c-t|\big)\Big];\\
		\partial\lppls / \partial A&=1;\\
		\partial\lppls / \partial B&=|t_c-t|^m;\\
		\partial\lppls / \partial C_1&=|t_c-t|^m\cos\big(\omega\ln|t_c-t|\big);\\
		\partial\lppls / \partial C_2&=|t_c-t|^m\sin\big(\omega\ln|t_c-t|\big).\\
	\end{array}
\end{equation}
The second-order derivatives $\partial^2 \lppls/\partial\psi_i\partial\psi_j$, which are needed for the calculation of the matrix $H$~\eqref{eq:H}, have the following form (omitting equivalent symmetrical entries, i.e.: $\partial^2\lppls/ \partial m\partial\omega\equiv\partial^2\lppls/ \partial \omega\partial m$):
\begin{equation}\label{eq:LPPLS_partial_2}
	\begin{array}{llll}
		\partial^2\lppls &/& \partial m^2&=|t_c-t|^m\left(\ln|t_c-t|\right)^2\Big[
		B+C_1\cos\big(\omega\ln|t_c-t|\big)+C_2\sin\big(\omega\ln|t_c-t|\big)\Big];\\
		\partial^2\lppls &/& \partial m\partial\omega&=|t_c-t|^m\left(\ln|t_c-t|\right)^2\Big[
		-C_1\sin\big(\omega\ln|t_c-t|\big)+C_2\cos\big(\omega\ln|t_c-t|\big)\Big];\\
		\partial^2\lppls &/& \partial m\partial B&=|t_c-t|^m\ln|t_c-t|;\\
		\partial^2\lppls &/& \partial m\partial C_1&=|t_c-t|^m\ln|t_c-t|\cos\big(\omega\ln|t_c-t|\big);\\
		\partial^2\lppls &/& \partial m\partial C_2&=|t_c-t|^m\ln|t_c-t|\sin\big(\omega\ln|t_c-t|\big);\\
		\partial^2\lppls &/& \partial\omega^2&=-|t_c-t|^m\left(\ln|t_c-t|\right)^2\Big[
		C_1\cos\big(\omega\ln|t_c-t|\big)+C_2\sin\big(\omega\ln|t_c-t|\big)\Big];\\
		\partial^2\lppls &/& \partial \omega\partial C_1&=-|t_c-t|^m\ln|t_c-t|\sin\big(\omega\ln|t_c-t|\big);\\
		\partial^2\lppls &/& \partial \omega\partial C_2&=|t_c-t|^m\ln|t_c-t|\cos\big(\omega\ln|t_c-t|\big).\\
	\end{array}
\end{equation}
All other second-order partial derivatives are equal to zero.

\section{Jacobian matrix for the damping parameter}\label{app:jacobian_damping}

The Jacobian matrix for the parameter transformation from $\eta=\{m,\omega,A,B,C_1,C_2,s\}$ to $\zeta=\{D,\omega,A,B,C_1,C_2,s\}$, where $D=m|B|/\omega|C|$ has the following form
\begin{equation}\label{eq:jacobian_D}
	J_D=\frac{\partial\eta}{\partial\zeta}=\begin{pmatrix}
	\frac{\omega|C|}{|B|} & \frac{D|C|}{|B|} &0 & -\frac{D\omega|C|}{B|B|} & 
	\frac{D\omega C_1}{|B||C|} & \frac{D\omega C_2}{|B||C|} & 0\\
	0& 1& 0& 0& 0& 0& 0\\
	0& 0& 1& 0& 0& 0& 0\\
	0& 0& 0& 1& 0& 0& 0\\
	0& 0& 0& 0& 1& 0& 0\\
	0& 0& 0& 0& 0& 1& 0\\
	0& 0& 0& 0& 0& 0& 1\\
	\end{pmatrix},
\end{equation}
where $|C|=\sqrt{C_1^2+C_2^2}$.

\section{Illustration of the differences between profile and modified profile likelihood intervals of nuisance parameters}\label{sec:double_omega}

\begin{figure}[h!]
  \centering
  \includegraphics[width=0.9\textwidth]{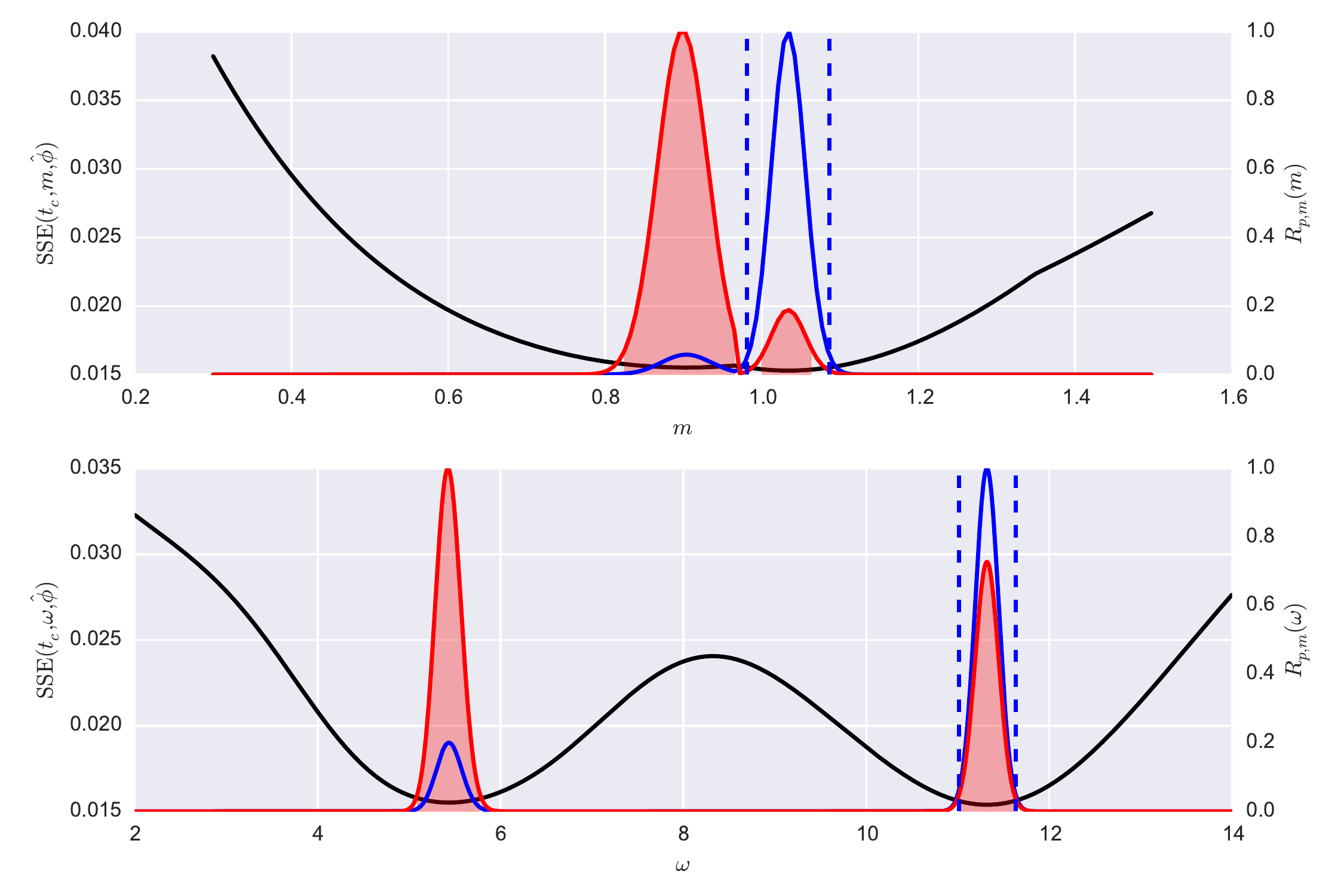}
  \caption{Profile of the cost function $F(\cdot)$ (black line, left scale), profile likelihood $L_p(\cdot)$ (blue line, right scale) and modified profile likelihood (red line, right scale) for the power law exponent $m$ (top subplot) and the logperiodic angular frequency $\omega$ (bottom subplot) for $t_c$=2007-11-20. The red shaded intervals show the likelihood intervals $\mathrm{LI}(\cdot)$ at the 5\% cutoff level. Vertical blue dashed lines denote approximated likelihood intervals~\eqref{eq:approx_LI_m_w} at the 5\% cutoff level. The calibration of the LPPLS model is performed on the Chinese SSEC Index for the bubble that bursts in June 2015.} 
  \label{fig:double_omega}
\end{figure}

Figure~\eqref{fig:double_omega} illustrates a situation in which the approximate likelihood intervals~\eqref{eq:approx_LI_m_w} are misleading. It presents the profile and modified profile likelihoods for the nuisance parameters $m$ and $\omega$ obtained by calibrating the LPPLS
model to the Chinese SSEC Index in the time window from $t_1$=2006-05-04, $t_2$=2007-10-31 and at the fixed $t_c$= 2007-11-20. In contrast to the typical situation shown in Figure~\ref{fig:profile_m_w}, one can clearly observe a bi-modal structure of the profile likelihoods of the nuisance parameters. Such bi-modal structure cannot be well described by intervals derived from a Fisher information-based likelihood. Moreover, this figure illustrates a case when the second-order modified profile likelihood suggests different estimated value of $m$ and $\omega$ compared with the standard MLE: profile and modified profile likelihood have maxima at different points (similarly to the situation of the critical time in Figure~\ref{fig:modified_profile}).

While these situations are rather rare according to our experience, one needs to be aware that the approximate relations~\eqref{eq:approx_LI_m_w} might not reflect the full complexity of the structure of residuals.


\section*{References}
\bibliographystyle{plainnat}
\bibliography{article_lppls_likelihood_v6}

\end{document}